\documentclass[a4paper,11pt]{article}

\usepackage[top=1in, bottom=1in, left=1in, right=1in]{geometry}

\usepackage{amsmath,amsthm,amssymb}
\usepackage{bm}
\usepackage[linesnumbered,ruled]{algorithm2e}
\usepackage{algpseudocode}
\usepackage{setspace}
\usepackage{siunitx}
\usepackage{csquotes}
\usepackage{multirow}
\usepackage{graphicx}
\usepackage{color}
\usepackage[caption=false, justification=centering]{subfig}

\usepackage{hyperref}
\usepackage{datetime}

\SetKwInOut{Input}{Input}
\SetKwInOut{Output}{Output}
\DeclareMathOperator{\Tr}{Tr}

\graphicspath{{figures/}}

\begin{document}

\title{A Linear Model for Microwave Imaging of Highly Conductive Scatterers}

\author{
        Shilong~Sun\thanks{S. Sun, B. J. Kooij, and A. G. Yarovoy are with the Delft University of Technology, 2628 Delft,
        The Netherlands (e-mail: shilongsun@icloud.com; B.J.Kooij@tudelft.nl; A.Yarovoy@tudelft.nl).}
        \and Bert~Jan~Kooij\footnotemark[1]
        \and Alexander~G.~Yarovoy\footnotemark[1]}

\newdate{date}{1}{11}{2017}
\date{\displaydate{date}}

\maketitle

\begin{abstract}

  In this paper, a linear model based on multiple measurement vectors model is proposed to formulate the inverse scattering problem of highly conductive objects at one single frequency. Considering the induced currents which are mostly distributed on the boundaries of the scatterers, joint sparse structure is enforced by a sum-of-norm regularization. Since no \textit{a priori} information is required and no approximation of the scattering model has been made, the proposed method is versatile. Imaging results with transverse magnetic and transverse electric polarized synthetic data and Fresnel data demonstrate its higher resolving ability than both linear sampling method and its improved version with higher, but acceptable, computational complexity.

\end{abstract}


\section{Introduction} \label{sec:intro}

    Inverse scattering is a procedure of recovering the characteristics of the objects from the scattered fields. It is of great importance because of its wide applications in different areas, such as seismic detection, medical imaging, sonar, remote sensing, and so forth. Most of the studies on the inverse scattering problems are focused on the frequencies of the resonant region, i.e., the wavelength is comparable to the dimension of the object. Challenges mainly lie in the nonlinearity and ill-posedness in the Hadamard sense \cite{hadamard2014lectures}. There is a large variety of possible inverse scattering problems, for example, find the shape of the scatterer with the boundary condition already known, or find the space dependent coefficients of the object without any \textit{a priori} information at all. The inverse scattering problem discussed in this paper is to determine the shape of the highly conductive scatterers with the scattered electromagnetic (EM) field for one or several incident fields at one single frequency of the resonant region. 

    Basically, there are two families of methods for solving this problem: the volume-based methods and the surface-based methods. The volume-based methods indicate the shape with space dependent coefficients. P. M. van den Berg proposed to retrieve the boundary of the highly conductive scatterer by doing the same with the iterative method of reconstructing the conductivity of an EM penetrable objects \cite{kleinman1994two}. This idea was further extended to the mixed dielectric and highly conductive objects combined with contrast source inversion (CSI) method \cite{van1997contrast}, see O F{\'e}ron et al. \cite{feron2005microwave}, Chun Yu et al. \cite{yu2005inversion}. Classical design of the cost functional consists of the data error and the state error without considering the cross-correlated mismatch of both errors. Recently, a so-called cross-correlated error function and a novel inversion method, referred to as cross-correlated CSI, have been proposed to enhance the inversion ability \cite{sun2016Cross}. The idea of solving the nonlinear inverse scattering problem with linear algebra can also be found in \cite{di2009numerical}. An algorithmically efficient algorithm, the time-reversal multiple signal classification (TR-MUSIC) method \cite{devaney2005time,lee2011compressive}, is also of interest since it is capable to break through the diffraction limit. Linear sampling method (LSM) \cite{colton1996simple,colton1997simple,crocco2013improved} is another typical volume-based method, which finds an indicator function for each voxel in the region of interest (ROI) by first defining a far-field mapping operator (or a near-field mapping operator \cite{fata2004linear}), then solving a linear system of equations. The norm of the indicator function approaches to zero when the position of the corresponding voxel approaches the highly conductive scatterer. Although LSM has been proved to be effective for highly conductive scatterer, and also applicable for dielectric scatterer in some cases \cite{arens2003linear}, this method needs sufficient amount of measurements to guarantee the inversion performance \cite{colton2013inverse}. Besides, it is very time-consuming to compute the dyadic Green function for an irregular inhomogeneous background \cite{eskandari2014three}, for instance, in the case of ground penetrating radar (GPR) \cite{daniels2005ground}. The surface-based methods first parameterize the shape of the scatterer mathematically with several parameters, then optimize the parameters by minimizing a cost function iteratively \cite{roger1981Newton}. The drawback of this method is obvious: 1) This type of methods require \textit{a priori} information on the position and quantity of the scatterers, more research on this point can be found in \cite{qing2003electromagnetic,qing2004electromagnetic}; 2) It is intractable to deal with the complicated nonconvex objects. Apart from that, each iteration involves finding a solution to a forward scattering problem, which is extremely time-consuming for the large-scale inverse problems with an irregular background. As a matter of fact, this is a general drawback of the iterative inversion methods. In cases where the dimension of the solution space is not so huge, global optimization techniques \cite{caorsi2001crack,rocca2009evolutionary,salucci2017multifrequency} are good candidates to search for the global optimal solution. We also refer to \cite{poli2013mt} for a compressive sensing CSI method which solves the contrast source two-step formulation for detecting the non-radiating part of the equivalent currents. 

    Recently we have proposed a linear model to address the nonlinear highly conductive inverse scattering problem with transverse magnetic (TM) polarized incident fields \cite{7571336}. The basic idea is to transfer the problem to a set of linear inverse source problems, formulate the set of problems with the multiple measurement vectors (MMV) model \cite{van2010theoretical}, and finally solve the problem with the sum-of-norm regularization constraint. We have also considered a cascade of the linear inverse source model and a linear optimization model for solving the 3-D inverse scattering problem with half-space configurations \cite{sun2016Linearied}. Although the feasibility of this idea has been demonstrated numerically, a theoretical framework has not been established yet. The extension in solving the inverse scattering problem of vectorial fields is not straightforward and therefore has to be proved theoretically. Moreover, the feasibility needs to be further demonstrated with experimental measurement data. In this paper, we first presented the theoretical framework in solving the highly conductive inverse scattering problem with TM-polarized incident fields. Based on the convex optimization theory and the spectral projected gradient method (SPGL1) \cite{BergFriedlander:2008,van2011sparse}, we extended the theoretical framework to solve the transverse electric (TE) polarized (vectorial field) inverse scattering problems. Consequently, the extension in 3-D problems can be derived straightforwardly. Cross-validation (CV) technique \cite{ward2009compressed,zhang2016cross} is used to terminate the iteration such that the estimation of noise level is well circumvented. Both 2-D synthetic data generated by a MATLAB-based ``MaxwellFDFD'' package \cite{W.Shin2013} and the experimental TM- and TE-polarized datasets of the Institut Fresnel, France \cite{0266-5611-17-6-301} are processed for demonstrating the validity of the proposed model. What's more, we have also presented an analysis on the computational complexity and the affect of transmitter/receiver number on the imaging quality of the proposed approach, which sheds a light on the design of the imaging system.

    For the case of penetrable objects, the contrast sources are distributed everywhere in the interior of the object. Since the linear model is regularized with sum-of-norm constraint, the reconstruction algorithm tends to seek for a solution of the minimum sum-of-norm. According to the field equivalence principle \cite{rengarajan2000field}, for penetrable object, there would more likely be several sparse solutions which not only generate the same scattered field pattern (satisfying the data equation) but also have the same non-zero structure (possessing smaller sum-of-norm than the real solutions). Since correct recovery cannot be guaranteed for penetrable objects, we restricted the discussion in this paper to highly conductive objects.

    In a nutshell, major differences of the proposed approach in comparison to other existing methods are as follows:
    \begin{enumerate}
      \item In comparison to linear methods with linear model of weak scattering assumptions (such as Born/Rytov approximations), the proposed model is more applicable since no weak scattering approximation has been made.
      \item In comparison to linear methods with linear model of no weak scattering approximations (such as linear sampling method), the joint sparse structure of the contrast sources is enforced in the proposed approach by use of sum-of-norm regularization constraint, resulting in higher resolving ability.
      \item In comparison to linear iterative algebra of multiple levels with non-linear model (such as CSI), calculation of the total fields for the proposed approach is not needed, resulting in higher imaging efficiency.
      \item In comparison to super resolution methods based on pseudo-spectrum analysis (such as Time-reversal MUSIC), the proposed approach does not need to estimate the scatterer number, nor does it need to care about how the imaging domain is discretized.
    \end{enumerate}
    Since the proposed approach is also based on a linear model of no weak scattering approximation, LSM and an improved version are selected in this paper for comparison.

    The remainder of the paper is organized as follows: In Subsection \ref{subsec.ProSta}, the problem statement is given; In Subsection \ref{subsec.For}, the formulation of the linear model is presented; In Subsection \ref{subsec.SMV_TM}, the SPGL1 method for solving the single measurement vector (SMV) model in TM case is introduced; In Subsection \ref{subsec.MMV_TM} and \ref{subsec.MMV_TE}, we derived the sum-of-norm optimization method for solving the MMV model\footnote{The multi-frequency implementation of this method (referred to as GMMV-LIM) is available at \url{https://github.com/TUDsun/GMMV-LIM}.} of TM case and TE case, respectively; In Subsection \ref{subsec.CVSPGL1}, a CV-based modified SPGL1 method is introduced; In Section \ref{sec.LSM}, linear sampling method and its improved version are introduced; The inverted results with synthetic data and experimental data are given in Section \ref{sec.Syn} and Section \ref{sec.Exp}, respectively; Finally, Section \ref{sec.Con} ends the paper with our conclusions.

\section{MMV Linear Inversion Model}

    \subsection{Problem statement}\label{subsec.ProSta}

\begin{figure}[!t]
            \centering
            \includegraphics[height=0.45\linewidth]{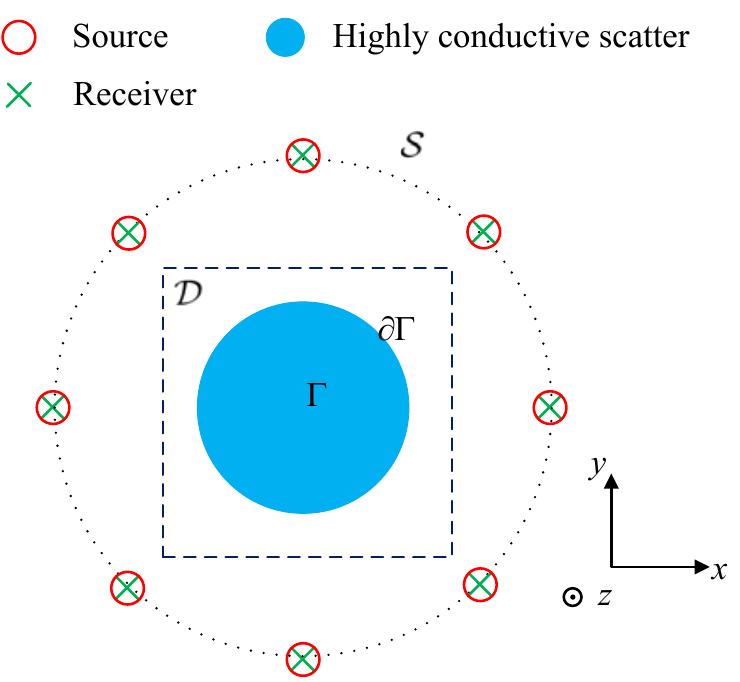}
            \caption{The configuration of the inverse scattering problem with respect to highly conductive scatterers.}
            \label{fig:probSta}
        \end{figure}

        We consider a scattering configuration as depicted in Figure~\ref{fig:probSta}, which consists of a bounded, simply connected, inhomogeneous background domain $\mathcal{D}$. The domain $\mathcal{D}$ contains a highly conductive cylinder $\Gamma$, whose surface is represented by $\partial\Gamma$. The dielectric properties of the background are known beforehand. The domain $\mathcal{S}$ contains the sources and receivers. The sources are denoted by the subscript $p$ (where $p\in\{1,2,3...,P\}$), and the receivers are denoted by the subscript $q$ (where $q\in\{1,2,3,...,Q\}$). Sources and receivers that have equal subscripts are located at the same position. We use a right-handed coordinate system in which the unit vector in the invariant direction points out of the paper. 

        In our notation for the vectorial quantities, we use a bold notation which represents a vector with three components. The general mathematical representations presented are consistent with any 3-D configuration, in which the 2-D TE and TM excitations are a special case, resulting in vectors containing zero elements. In this paper, we consider the time factor $\exp(\text{i}\omega t)$. Here, $\text{i}$ represents the imaginary unit, then the vectorial Maxwell's equations in the frequency domain can be written as
        \begin{equation}\label{eq.maxwell}
        \begin{array}{ccc}
          \begin{bmatrix}
          -\text{i} \omega\bm{\varepsilon}-\bm{\sigma} & \nabla\times \\
          \nabla\times & \text{i}\omega \bm{\mu} \\
        \end{bmatrix} 
        & 
        \begin{bmatrix}
          \bm{E}\\ 
          \bm{H}
        \end{bmatrix}
        &= 
        \begin{bmatrix}
          \bm{J}\\
          -\bm{M}
        \end{bmatrix}\\
        \end{array},
        \end{equation}
        where $\bm{E}$ and $\bm{H}$ are the electric and magnetic fields, respectively; $\bm{J}$ and $\bm{M}$ are the electric and magnetic current source densities, respectively; $\bm{\sigma}$, $\bm{\varepsilon}$ and $\bm{\mu}$ are the electric conductivity, electric permittivity and magnetic permeability, respectively. For most of the real problems, $\bm{\mu}$ can be reasonably assumed to be the permeability of free space, $\mu_0$, while $\bm{\sigma}$ and $\bm{\varepsilon}$ are functions of both the position vector $\bm{x}=[x_1,x_2,x_3]^T$ and the angular frequency $\omega$.

        Since the solution of the $E$-field is two orders of magnitude larger than the $H$-field, it is for better numerical accuracy (see \cite{W.Shin2013}) to eliminate either the $E$-field or the $H$-field from equation~\eqref{eq.maxwell}. In this paper, we assume the electric field is measured, so we rewrite our equations in terms of the electric field $\bm{E}$ according to
        \begin{equation}\label{eq.E}
            \nabla\times\bm{\mu}^{-1}\nabla\times\bm{E}-\omega^2\bm{\epsilon}\bm{E}=-\text{i}\omega\bm{J}-\nabla\times\bm{\mu}^{-1}\bm{M},
        \end{equation}
        where $\bm{\epsilon}$ is the complex permittivity given by 
        \begin{equation}
            \bm{\epsilon}=\bm{\varepsilon}-\text{i}\bm{\sigma}/\omega.
        \end{equation}

        The total electric field, $\bm{E}_p$, and the incident electric field, $\bm{E}_p^{\text{inc}}$, are excited by the $p$-th external source, the scattered electric field $\bm{E}_p^{{\text{sct}}}$ is then found by
        \begin{equation}\label{eq.sct.inc}
            \bm{E}_p^{{\text{sct}}}=\bm{E}_p-\bm{E}_p^{\text{inc}},\quad p=1,2,\dots,P.
        \end{equation}
        Here, the subscript $p$ corresponds to the $p$-th source. According to the above relation and the electric field equation \eqref{eq.E}, it is easy to obtain the basic equation of the inverse scattering problem, which is
        \begin{equation}\label{eq.CSI.E}
            \nabla\times\mu_0^{-1}\nabla\times\bm{E}_p^{{\text{sct}}}-\omega^2\bm{\epsilon}_{\text{b}}\bm{E}_p^{{\text{sct}}} = \omega^2\bm{\chi}\bm{E}_p,\quad p=1,2,\dots,P,
        \end{equation} 
        where, the contrast $\bm{\chi}$ is the difference of the complex permittivity of the test domain, $\bm{\epsilon}$, and that of the background, $\bm{\epsilon}_{\text{b}}$, i.e., $\bm{\chi}=\bm{\epsilon}-\bm{\epsilon}_{\text{b}}$. The problem we are going to resolve is to find the shape of the highly conductive scatterer $\partial\Gamma$ from the measurement of the scattered electric fields $\bm{E}_p^{{\text{sct}}}$. Since the total electric field $\bm{E}_p$ is a function of the contrast $\bm{\chi}$, this is obviously a nonlinear problem. 

    \subsection{Formulation}\label{subsec.For}

        First, let us formulate the inverse problem following the vector form of the FDFD scheme in \cite{W.Shin2013}, and rewrite \eqref{eq.CSI.E} as follows 
        \begin{equation}\label{eq.sct.E.eq}
            \bm{A}\bm{e}_p^{\text{sct}}=\omega^2 \bm{\chi} \bm{e}_p,\quad p = 1,2,3,\cdots,P,
        \end{equation}
        where, $\bm{A}$ is the stiffness matrix, and 
        \begin{equation}
            \bm{e}_p^{\text{sct}}=\begin{bmatrix}
                \bm{e}_{p,x_1}^{\text{sct}}\\
                \bm{e}_{p,x_2}^{\text{sct}}\\
                \bm{e}_{p,x_3}^{\text{sct}}
            \end{bmatrix},\quad
            \bm{e}_p = 
            \begin{bmatrix}
                \bm{e}_{p,x_1}\\
                \bm{e}_{p,x_2}\\
                \bm{e}_{p,x_3}
            \end{bmatrix},
        \end{equation}
        and $\bm{\chi} \bm{e}_p$ is the component-wise multiplication of the two vectors, $\bm{\chi}$ and $\bm{e}_p$. The scattered fields $\bm{e}^{\text{sct}}_p=\bm{A}^{-1}\omega^2\bm{\chi}\bm{e}_p$ are probed and the measurements are used to estimate the unknown $\bm{\chi}$. Now let us use a measurement operator, $\mathcal{M}_{\mathcal{S}}$, to select the field values at the receiver positions, then we can formulate the data equations as $\bm{y}_p=\mathcal{M}_{\mathcal{S}}\bm{A}^{-1}\omega^2\bm{\chi}\bm{e}_p$, where $\bm{y}_p$ is the measurement vector of the $p$-th scattered fields. Let us further define the contrast source as $\bm{j}_p:=\bm{\chi}\bm{e}_p$, then we have
        \begin{equation}\label{eq.FD-CSI.eq}
            \bm{y}_p = \bm{\Phi} \bm{j}_p,\quad p = 1,2,3,\cdots,P,
        \end{equation}
        where, $\bm{\Phi}$ is the sensing matrix defined by
        \begin{equation}
            \bm{\Phi} = \mathcal{M}_{\mathcal{S}}\bm{A}^{-1}\omega^2.
        \end{equation}
        In TM case, $\bm{\Phi}\in\mathbb{C}^{Q\times N}$, while in TE case, $\bm{\Phi}\in\mathbb{C}^{2Q\times 2N}$. Here, $N$ is the grid number of the discretized inversion domain.

        Since the contrast source $\bm{j}_p$ shows sparsity due to the fact that the induced current only exists on the surface of the highly conductive objects, the ill-posedness of the inverse scattering problem can be overcome by exploiting the sparsity of the contrast sources. Further, although the contrast sources $\bm{j}_p$ excited by the illumination of the incident fields $\bm{e}^\text{inc}_p$ are of different values, the non-zero elements are located at the same positions --- the boundary of the highly conductive scatterers. This inspired us to improve the inversion performance by enhancing the joint sparse structure, so the linear data model \eqref{eq.FD-CSI.eq} is further formulated as an multiple measurement vectors model. In the following part of this paper, a linear model is constructed and a sum-of-norm optimization problem is derived for TM and TE cases, respectively. In doing so, the nonlinear inverse scattering problem can be simplified and addressed by a linear optimization scheme without considering the state equations (i.e., the calculation of the total fields is circumvented)
        \begin{equation}\label{eq.state}
            \bm{e}_p = \bm{e}^\text{sct}_p + \bm{A}^{-1}\omega^2 \bm{j}_p,\quad  \quad p=1,2,\dots,P.
        \end{equation} 
        Specifically, Eq. \eqref{eq.FD-CSI.eq} is rewritten as follows
        \begin{equation}
            \bm{Y} = \bm{\Phi} \bm{J}+\bm{U},
        \end{equation}  
        where, $\bm{\Phi}$ is the joint sensing matrix, \[\bm{J}=\left[\bm{j}_1,\bm{j}_2,\cdots,\bm{j}_P\right]\] is the contrast source matrix, and \[\bm{Y}=\left[\bm{y}_1,\bm{y}_2,\cdots,\bm{y}_P\right]\] is the measurement data matrix, and $\bm{U}$ represents the additive complex measurement noise matrix.

    \subsection{Solving the SMV model: TM case}\label{subsec.SMV_TM}

        First, consider the single source configuration illuminated by TM-polarized wave. The inverse scattering problem is formulated as a basis pursuit denoise (BP$_{\widetilde\sigma}$) problem \cite{sun2015novel}:
        \begin{equation}
            (\text{SMV}_{\text{TM}}\ \text{BP}_{\widetilde\sigma}): \ \text{min}\ \left\|\bm{j}_p\right\|_1\  \text{s.t.}\   \left\|\bm{\Phi}\bm{j}_p - \bm{y}_p\right\|_2\leq \widetilde\sigma.
        \end{equation}
        where, $\widetilde\sigma$ represents the noise level, and the contrast source is regularized with the $\ell_1$-norm constraint. Solving this problem means searching for a solution $\bm{j}_p$ which is of the smallest $\ell_1$-norm and meanwhile satisfies the inequality condition. Although the BP$_{\widetilde\sigma}$ problem is straightforward for understanding the inverse problem, it is not easy to solve directly even if we exactly know the value of $\widetilde\sigma$. An equivalent problem that is much simpler to solve is the Lasso (LS$_\tau$) problem \cite{tibshirani1996regression}, which is formulated as 
        \begin{equation}\label{eq.LS}
            (\text{SMV}_{\text{TM}}\ \text{LS}_\tau): \ \text{min}\ \left\|\bm{\Phi}\bm{j}_p - \bm{y}_p\right\|_2\  \text{s.t.}\  \left\|\bm{j}_p\right\|_1 \leq \tau.
        \end{equation}
        The LS$_\tau$ problem can be solved using a spectral projected gradient (SPG) method that is proposed based on convex optimization theory \cite{birgin2000nonmonotone,birgin2003inexact,dai2005projected}. Details of the SPG method for solving the (LS$_\tau$) problem \eqref{eq.LS} is given by \cite[Algorithm 1]{BergFriedlander:2008}, in which $\mathcal{P}_\tau[\cdot]$ is a projection operator defined as
        \begin{equation}
            \mathcal{P}_\tau\left[\bm{j}_p\right]:=\left\{\underset{\bm{s}}{\arg \min}\quad \left\|\bm{j}_p-\bm{s}\right\|_2 \ \text{s.t.}\ \|\bm{s}\|_1 \leq \tau \right\}.
        \end{equation}
        $\mathcal{P}_\tau[\cdot]$ gives the projection of a vector $\bm{j}_p$ onto the one-norm ball with radius $\tau$.

        In practice, $\tau$ is usually not available. For solving this problem, a Pareto curve is defined in SPGL1 algorithm \cite{BergFriedlander:2008} by
        \begin{equation}
            \phi_{\text{SMV}_{\text{TM}}}(\tau) = \left\|\bm{\Phi}\bm{j}_{p,\tau} - \bm{y}_p\right\|_2, 
        \end{equation}
        where, $\bm{j}_{p,\tau}$ is the optimal solution to the LS$_\tau$ problem. It is easy to find that the (BP$_{\widetilde\sigma}$) problem is equivalent to the (LS$_\tau$) problem when $\phi_{\text{SMV}_{\text{TM}}}(\tau)=\widetilde\sigma$ is satisfied.
        \begin{figure}[!t]
            \centering
            \includegraphics[height=0.45\linewidth]{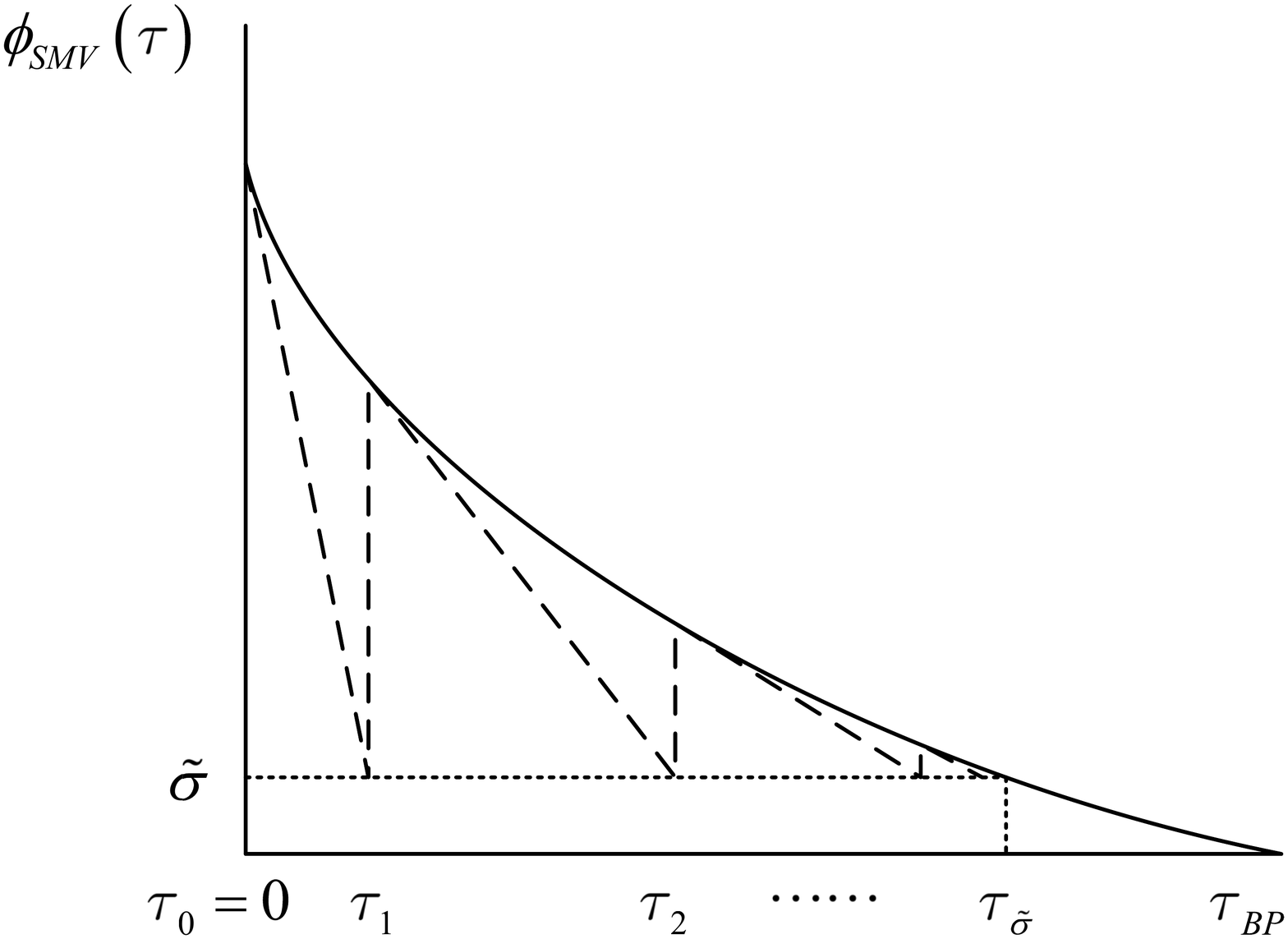}
            \caption{Probing the Pareto curve: the update of parameter $\tau$.}
            \label{fig:Pareto}
        \end{figure}
        The Pareto curve is proved to be differentiable under some conditions, and the root of the nonlinear equation $\phi_{\text{SMV}_{\text{TM}}}(\tau)=\widetilde\sigma$ can be reached by Newton iterations \cite{BergFriedlander:2008}
        \begin{equation}
            \tau_{h+1} = \tau_{h}+\frac{\widetilde\sigma-\phi_{\text{SMV}_{\text{TM}}}\left(\tau_{h}\right)}{\phi_{\text{SMV}_{\text{TM}}}'\left(\tau_{h}\right)},
        \end{equation} 
        where,
        \begin{equation}
            \phi'_{\text{SMV}_{\text{TM}}}(\tau_{h+1}) = -\frac{\left\|\bm{\Phi}^H\bm{r}_{p,\tau_h}\right\|_{\infty}}{\|\bm{r}_{p,\tau}\|_2}.
        \end{equation}
        where, $\bm{\Phi}^H$ is the conjugate transpose of matrix $\bm{\Phi}$, $\bm{r}_{p,\tau_h}=\bm{\Phi}\bm{j}_{p,\tau_h} - \bm{y}_p$ is the residual vector. The update of $\tau$ by probing the Pareto curve is illustrated in Figure \ref{fig:Pareto}. This procedure requires computing successively more accurate solutions of LS$_\tau$. The Newton root-finding framework for solving the ($\text{SMV}_{\text{TM}}\ \text{BP}_{\widetilde\sigma}$) problem is given in \cite[Algorithm 1]{van2011sparse}.

    \subsection{Solving the MMV model: TM case}\label{subsec.MMV_TM}

        Let us now first consider the 2-D multi-source configuration with TM-polarized illumination. As all the contrast sources are focused on the boundary of the scatterers, the contrast source matrix $\bm{J}$ shows row-sparsity. Therefore, the inverse scattering problem with multi-source configurations can be formulated as an ($\text{MMV}_{\text{TM}}\ \text{BP}_{\widetilde\sigma}$) problem regularized by a sum-of-norm constraint 
        \begin{equation}
            (\text{MMV}_{\text{TM}}\ \text{BP}_{\widetilde\sigma}) \ \text{min}\ \kappa(\bm{J})\  \text{s.t.}\  \|\bm{\Phi}\bm{J} - \bm{Y}\|_F\leq \widetilde\sigma,
        \end{equation}
        where, $\kappa(\bm{J})$ is the mixed ($\alpha,\beta$)-norm defined as 
        \begin{equation}
            \|\bm{J}\|_{\alpha,\beta}:=\left(\sum_{i=1}^N\left\|\bm{J}_{i,:}^T\right\|_\beta^\alpha\right)^{1/\alpha},
        \end{equation}
        with $\bm{J}_{i,:}$ denoting the $i$-th row of $\bm{J}$, and $\|\cdot\|_\beta$ the conventional $\beta$-norm. $\|\cdot\|_F$ is the Frobenius norm which is equivalent to the mixed (2,2)-norm $\|\cdot\|_{2,2}$. In this problem, we consider $\alpha=1$ and $\beta=2$, which is a sum-of-norm constraint. Accordingly, the $\text{MMV}_{\text{TM}}\ \text{LS}_\tau$ problem is reformulated as
        \begin{equation}\label{eq.LSMMV}
            (\text{MMV}_{\text{TM}}\ \text{LS}_{\tau}) \ \text{min}\ \left\|\bm{\Phi}\bm{J} - \bm{Y}\right\|_F\  \text{s.t.}\  \left\|\bm{J}\right\|_{1,2} \leq \tau,
        \end{equation}
        and the Pareto curve for the MMV model is defined as
        \begin{equation}\label{eq.Pareto_MMVTM}
            \phi_{\text{MMV}_{\text{TM}}}(\tau) = \left\|\bm{\Phi}\bm{J}_{\tau} - \bm{Y}\right\|_F,
        \end{equation}
        where, $\bm{J}_{\tau}$ is the optimal solution to the LS$_\tau$ problem \eqref{eq.LSMMV}.

        \begin{algorithm}[!t]\setstretch{1.0}
            \caption{Spectral projected gradient for ($\text{MMV}_{\text{TM}}\ \text{LS}_\tau$) problem.}\label{alg.SPGTM}
            \Input{$\bm{\Phi}$, $\bm{Y}$, $\bm{J}$, $\tau$}
            \Output{$\bm{J}_{\tau}$, $\bm{R}_{\tau}$}
            Set minimum and maximum step lengths $0<\alpha_{\min}<\alpha_{\max}$\;
            Set initial step length $\alpha_0\in[\alpha_{\min},\alpha_{\max}]$ and sufficient descent parameter $\gamma\in(0,0.5)$\;
            Set an integer line search history length $M\geq 1$\;
            $\bm{J}_{0}=\mathcal{P}_{\tau,\text{MMV}_{\text{TM}}}\left[\bm{J}\right]$, $\bm{R}_{0}=\bm{Y}-\bm{\Phi}\bm{J}_{0}$, $\bm{G}_{0}=-\bm{\Phi}^H\bm{R}_{0}$, $\ell = 0$\;
            \Begin
              {
                $\delta_\ell \gets \left|\left\|\bm{R}_{\ell}\right\|_F-\frac{\textcolor{red}{\Tr}\left\{\bm{Y}^H\bm{R}_{\ell}\right\}-\tau\|\bm{G}_{\ell}\|_{\infty,2}}{\left\|\bm{R}_{\ell}\right\|_F}\right|$ \Comment{compute duality gap}\;
                \textbf{If} $\delta_\ell \leq \epsilon$, \textbf{then} break\;
                $\alpha \gets \alpha_\ell$ \Comment{initial step length}\;
                \Begin
                {
                  $\overline{\bm{J}} \gets \mathcal{P}_{\tau,\text{MMV}_{\text{TM}}}\left[\bm{J}_{\ell}-\alpha\bm{G}_{\ell}\right]$ \Comment{projection}\;
                  $\overline{\bm{R}} \gets \bm{Y} - \bm{\Phi}\overline{\bm{J}}$ \Comment{update the corresponding residual}\;
                  \eIf{$\left\|\overline{\bm{R}}\right\|_F^2\leq \underset{h\in[0,\min\{\ell,M-1\}]}{\max}\left\|\bm{R}_{\ell-h}\right\|_F^2 + 
                  \gamma\Re\left\{\textcolor{red}{\Tr}\left\{\left(\overline{\bm{J}}-\bm{J}_{\ell}\right)^H\bm{G}_{p,\ell}\right\}\right\}$}
                  {
                    break\;
                  }
                  {
                    $\alpha\gets\alpha/2$\;
                  }
                }
                $\bm{J}_{\ell+1}\gets\overline{\bm{J}}$, $\bm{R}_{\ell+1}\gets \overline{\bm{R}}$, $\bm{G}_{\ell+1}\gets -\bm{\Phi}^H\bm{R}_{\ell+1}$ \Comment{update iterates}\; 
                $\Delta\bm{J}\gets\bm{J}_{\ell+1}-\bm{J}_{\ell}$, $\Delta \bm{G}\gets \bm{G}_{\ell+1}-\bm{G}_{\ell}$\;
                \eIf{$\Re\left\{\textcolor{red}{\Tr}\left\{\Delta\bm{J}^H\Delta\bm{G}\right\}\right\}\leq 0$} 
                {
                    $\alpha_{\ell+1}\gets\alpha_{\max}$ \Comment{update the Barzilai-Borwein step length}\;
                }
                {
                    $\alpha_{\ell+1}\gets\min\left\{\alpha_{\max},\max\left[\alpha_{\min},\frac{\textcolor{red}{\Tr}\left\{\Delta\bm{J}^H\Delta\bm{J}\right\}}{\Re\left\{\textcolor{red}{\Tr}\left\{\Delta\bm{J}^H\Delta\bm{G}\right\}\right\}}\right]\right\}$\;
                }
                $\ell\gets\ell+1$\;
              }
              \textbf{return} $\bm{J}_{\tau}\gets \bm{J}_{\ell}$, $\bm{R}_{\tau}\gets\bm{R}_{\ell}$\;
        \end{algorithm}

        According to \cite[Theorem 2.2]{van2011sparse} and \cite[Chapter 5]{van2009convex}, $\phi_{\text{MMV}_{\text{TM}}}(\tau)$ is continuously differentiable and 
        \begin{equation}\label{eq.PPareto_MMVTM}
            \phi'_{\text{MMV}_{\text{TM}}}(\tau_h) = -\frac{\left\|\bm{\Phi}^H(\bm{\Phi}\bm{J}_{\tau_h}-\bm{Y})\right\|_{\infty,2}}{\left\|\bm{\Phi}\bm{J}_{\tau_h}-\bm{Y}\right\|_F},
        \end{equation}
        where, $\|\cdot\|_{\infty,2}$ is the dual norm of $\|\cdot\|_{1,2}$ \cite[Corollary 6.2]{van2011sparse}. Similarly, the root of the nonlinear equation $\phi_{\text{MMV}_{\text{TM}}}(\tau)=\widetilde\sigma$ can also be reached by Newton iterations
        \begin{equation}
            \tau_{h+1} = \tau_{h}+\frac{\widetilde\sigma-\phi_{\text{MMV}_{\text{TM}}}\left(\tau_{h}\right)}{\phi_{\text{MMV}_{\text{TM}}}'\left(\tau_{h}\right)}.
        \end{equation} 
        The projection operator $\mathcal{P}_{\tau}[\cdot]$ is replaced by an orthogonal projection onto $\|\cdot\|_{1,2}$ balls, $\mathcal{P}_{\tau,\text{MMV}_{\text{TM}}}[\cdot]$, which is defined as follows
        \begin{equation}\label{eq.Projection_MMVTM}
            \mathcal{P}_{\tau,\text{MMV}_{\text{TM}}}[\bm{J}]:=\left\{\underset{\bm{S}}{\arg \min}\quad \left\|\bm{J} - \bm{S}\right\|_F \ \text{s.t.}\ \|\bm{S}\|_{1,2} \leq \tau \right\}.     
        \end{equation} 
        We refer to \cite[Theorem 6.3]{van2011sparse} for the implementation of the projection operator. The (MMV$_{\text{TM}}$ BP$_{\widetilde\sigma}$) problem is solved by Algorithm \ref{alg.SPGTM} and Algorithm \ref{alg.NewtonRootTM} with the Pareto curve, $\phi_{\text{MMV}_{\text{TM}}}(\tau)$, its derivative with respective to $\tau$, $\phi'_{\text{MMV}_{\text{TM}}}(\tau)$, and the projection operator, $\mathcal{P}_{\tau,\text{MMV}_{\text{TM}}}[\cdot]$, defined by \eqref{eq.Pareto_MMVTM}, \eqref{eq.PPareto_MMVTM}, and \eqref{eq.Projection_MMVTM}, respectively. 

        \begin{algorithm}[!t]\setstretch{1.0}
            \caption{Newton root-finding framework.}\label{alg.NewtonRootTM}
            \Input{$\bm{\Phi}$, $\bm{Y}$, $\widetilde\sigma$}
            \Output{$\bm{J}_{\widetilde\sigma}$}
            $\bm{J}_{0} \gets \bm{0}$, $\bm{R}_{0} \gets \bm{Y}$, $\tau_0 \gets 0$, $h\gets 0$\; 
            \Begin
            {
                \textbf{If} $\left|\|\bm{R}_h\|_F-\widetilde\sigma\right| \leq \epsilon$, \textbf{then} break\;
                Solve the ($\text{MMV}_{\text{TM}}\ \text{LS}_\tau$) problem for $\tau_h$ using Algorithm \ref{alg.SPGTM}\;
                $\bm{R}_h\gets \bm{\Phi}\bm{J}_{h}-\bm{Y}$\;
                $\tau_{h+1} \gets \tau_h + \frac{\widetilde\sigma - \phi_{\text{MMV}_{\text{TM}}}(\tau_h)}{\phi_{\text{MMV}_{\text{TM}}}'(\tau_h)}$ \Comment{Newton update}\;
                $h \gets h+1$\;
            }
              \textbf{return} $\bm{J}_{\widetilde\sigma}\gets \bm{J}_{h}$\;
        \end{algorithm}

    \subsection{Solving the MMV model: TE case}\label{subsec.MMV_TE}

        For the TE polarization case, the electric field is not a scalar anymore. Therefore, care must be given to the formulation of the (MMV$_{\text{TE}}$ BP$_{\widetilde\sigma}$) problem. Considering the two components of electric field, $E_x$ and $E_y$, the inverse scattering problem for the TE case can be formulated as
        \begin{equation}\label{MMVTEBP}
            (\text{MMV}_{\text{TE}}\ \text{BP}_{\widetilde\sigma}) \ \text{min}\ \kappa_{\text{TE}}(\bm{J}) \  \text{s.t.}\  \rho\left(\bm{\Phi}\bm{J} - \bm{Y}\right) \leq \widetilde\sigma,
        \end{equation}
        where, 
        \begin{equation}\label{eq.kappa}
            \kappa_{\text{TE}}(\bm{J}) := \sum_{n=1}^N\left\|\left[\bm{J}_{2n-1,:}\ \bm{J}_{2n,:}\right]^T\right\|_2,
        \end{equation}
        and
        \begin{equation}
            \rho(\cdot) := \|\cdot\|_F,
        \end{equation}
        are gauge functions \cite{rockafellar1970convex}. The $\text{MMV}_{\text{TE}}\ \text{LS}_\tau$ problem is formulated accordingly as
        \begin{equation}\label{MMVTELS}
            (\text{MMV}_{\text{TE}}\ \text{LS}_{\tau}) \ \text{min} \  \quad \rho\left(\bm{\Phi}\bm{J} - \bm{Y}\right) \  \text{s.t.} \  \kappa_{\text{TE}}(\bm{J}) \leq \tau.
        \end{equation}

        \subsubsection{Derivation of the dual}

        Let us rewrite \eqref{MMVTELS} in terms of $\bm{J}$ and an explicit residual term $\bm{R}$
        \begin{equation}
            \underset{\bm{J},\bm{R}}{\text{min}} \ \rho(\bm{R}) \ \text{s.t.} \ \bm{\Phi}\bm{J}+\bm{R} = \bm{Y},\quad \kappa_{\text{TE}}(\bm{J})\leq \tau.
        \end{equation}
        The dual to this equivalent problem is given by \cite[Chapter 5]{boyd2004convex}
        \begin{equation}
            \underset{\bm{Z},\lambda}{\text{max}} \quad \mathcal{G}(\bm{Z},\lambda) \ \text{s.t.} \ \lambda \geq 0,
        \end{equation}
        where $\bm{Z}\in\mathbb{C}^{(2M)\times P}$ and $\lambda\in\mathbb{C}$ are dual variables, and $\mathcal{G}$ is the Lagrange dual function, given by
        \begin{equation}
          \begin{split}
            \mathcal{G}(\bm{Z},\lambda):=&\underset{\bm{J},\bm{R}}{\inf}\left\{\rho(\bm{R})-\text{Tr}\left\{\bm{Z}^H(\bm{\Phi}\bm{J}+\bm{R}-\bm{Y})\right\} + \lambda\left(\kappa_{\text{TE}}(\bm{J})-\tau\right)\right\},
          \end{split}
        \end{equation}
        where $\text{Tr}$ represents the trace of a matrix. By separability of the infimum over $\bm{J}$ and $\bm{R}$ we can rewrite $\mathcal{G}$ in terms of two separate suprema, 
        \begin{equation}\label{eq.dual_sup}
          \begin{split}
            \mathcal{G}(\bm{Z},\lambda) = &\text{Tr}\left\{\bm{Y}^H\bm{Z}\right\}-\tau\lambda-\underset{\bm{R}}{\sup}\left\{\text{Tr}\left\{\bm{Z}^H\bm{R}\right\}-\rho(\bm{R})\right\}\\
            &-\underset{\bm{J}}{\sup}\left\{\text{Tr}\left\{\bm{Z}^H(\bm{\Phi}\bm{J})\right\}-\lambda\kappa_{\text{TE}}(\bm{J})\right\}
          \end{split}
        \end{equation}
        It is easy to see that the first supremum is the conjugate function of $\rho$ and the second supremum is the conjugate function of $\kappa_{\text{TE}}$ \cite[Chapter 3.3]{boyd2004convex}, by noting that
        \begin{equation}
          \begin{split}
            \text{Tr}\left\{\bm{Z}^H\bm{R}\right\} &= \text{vec}\{\bm{Z}\}^H\text{vec}\{\bm{R}\},\\
            \rho(\bm{R}) &= \rho(\text{vec}\{\bm{R}\}),
          \end{split}
        \end{equation}
        and 
        \begin{equation}
          \begin{split}
            \text{Tr}\left\{\bm{Z}^H(\bm{\Phi}\bm{J})\right\} &= \text{vec}\left\{\widetilde{\bm{Z}}\right\}^H\text{vec}\{\bm{J}\}, \\
            \kappa_{\text{TE}}(\bm{J}) &= \kappa_{\text{TE}}(\text{vec}\{\bm{J}\}),
          \end{split}
        \end{equation}
        respectively. Here, $\text{vec}\{\cdot\}$ is the vectorization of a matrix, $\widetilde{\bm{Z}}=\bm{\Phi}^H\bm{Z}\in\mathbb{C}^{(2N)\times P}$, and $\kappa_{\text{TE}}\left(\text{vec}\{\bm{J}\}\right)$ is defined equivalently as $\kappa_{\text{TE}}(\bm{J})$ in \eqref{eq.kappa}. Therefore, we have
        \begin{equation}\label{eq.conjugate_rho}
            \text{Tr}\left\{\bm{Z}^H\bm{R}\right\}-\rho(\bm{R})=
            \begin{cases}
                0\quad \rho^o(\bm{Z})\leq 1\\
                \infty\quad \text{otherwise}
            \end{cases},
        \end{equation}
        and
        \begin{equation}\label{eq.conjugate_kappa}
            \text{Tr}\left\{\bm{Z}^H(\bm{\Phi}\bm{J})\right\}-\lambda\kappa_{\text{TE}}(\bm{J}) = 
            \begin{cases}
                0\quad \kappa_{\text{TE}}^o(\widetilde{\bm{Z}})\leq \lambda\\
                \infty\quad \text{otherwise}
            \end{cases},
        \end{equation}
        where, the polar of $\rho$ and $\kappa_{\text{TE}}$ are defined by
        \begin{equation}
            \rho^o(\bm{Z}):=\underset{\bm{R}}{\sup}\left\{\left.\text{Tr}\left\{\bm{Z}^H\bm{R}\right\}\right|\rho(\bm{R}) \leq 1\right\},
        \end{equation}
        and 
        \begin{equation}
            \kappa_{\text{TE}}^o(\widetilde{\bm{Z}}):=\underset{\bm{J}}{\sup}\left\{\left.\text{Tr}\left\{\bm{Z}^H(\bm{\Phi}\bm{J})\right\}\right|\kappa_{\text{TE}}(\bm{J}) \leq \lambda\right\},
        \end{equation}
        respectively. If the gauge function is a norm, the polar reduces to the dual norm \cite[Section 3.3.1]{boyd2004convex}, i.e., $\rho^o(\bm{Z})=\|\bm{Z}\|_F$ and 
        \begin{equation}
          \begin{split}
            \kappa_{\text{TE}}^o(\widetilde{\bm{Z}}) =& \left(\sum_{n=1}^N\left\|\left[\widetilde{\bm{Z}}_{2n-1,:}\ \widetilde{\bm{Z}}_{2n,:}\right]\right\|_2^{\infty}\right)^{1/\infty}\\
            =& \max\left\{\left.\left\|\left[\widetilde{\bm{Z}}_{2n-1,:}\ \widetilde{\bm{Z}}_{2n,:}\right]\right\|_2\right|n=1,2,\cdots,N\right\}
          \end{split}
        \end{equation}
        (for more details see \cite[Corollary 6.2]{van2011sparse}). Substitution of \eqref{eq.conjugate_rho} and \eqref{eq.conjugate_kappa} into \eqref{eq.dual_sup} yields
        \begin{equation}\label{eq.dual}
            \underset{\bm{Z},\lambda}{\text{max}}\quad \text{Tr}\left\{\bm{Y}^H\bm{Z}\right\}-\tau\lambda\ \text{s.t.} \ \rho^o(\bm{Z})\leq 1,\quad \kappa_{\text{TE}}^o(\widetilde{\bm{Z}})\leq \lambda.
        \end{equation}

        In the case $\rho(\cdot)=\|\cdot\|_F$, the dual variable $\bm{Z}$ can be easily derived from
        \begin{equation}
            \underset{\bm{R}}{\sup}\quad \text{Tr}\left\{\bm{Z}^H\bm{R}\right\}-\|\bm{R}\|_F=0,\quad \text{if}\ \|\bm{Z}\|_F\leq 1,
        \end{equation}
        which is $\bm{Z}=\frac{\bm{R}}{\|\bm{R}\|_F}$. To derive the optimal $\lambda$, we can observe from \eqref{eq.dual} that as long as $\tau>0$, $\lambda$ must be at its lower bound $\kappa_{\text{TE}}^o(\widetilde{\bm{Z}})$, otherwise one can increase the objective $\text{Tr}\left\{\bm{Y}^H\bm{Z}\right\}-\tau\lambda$. Therefore, we obtain
        \begin{equation}
            \lambda = \frac{\kappa_{\text{TE}}^o\left(\bm{\Phi}^H\bm{R}\right)}{\|\bm{R}\|_F}.
        \end{equation}
        According to \cite[Theorem 5.2]{van2009convex}, we know that, on the open interval $\tau\in(0,\tau_0)$, where \[\tau_0=\min\left\{\tau\geq 0\left|\phi_{\text{MMV}_{\text{TE}}}(\tau)=\underset{\bm{J}}{\min}\ \rho(\bm{R})\right.\right\},\] the Pareto curve $\phi_{\text{MMV}_{\text{TE}}}(\tau)=\rho(\bm{R})$ is strictly decreasing, and continuously differentiable with 
        \begin{equation}
            \phi'_{\text{MMV}_{\text{TE}}}(\tau)=-\lambda=-\frac{\kappa_{\text{TE}}^o\left(\bm{\Phi}^H\bm{R}\right)}{\|\bm{R}\|_F}.
        \end{equation}
        The projection operator $\mathcal{P}_{\tau,\text{MMV}_{\text{TM}}}[\cdot]$ is replaced by an orthogonal projection onto $\kappa_{\text{TE}}(\cdot)$ balls, $\mathcal{P}_{\tau,\text{MMV}_{\text{TE}}}[\cdot]$, which is defined as follows
        \begin{equation}\label{eq.projectionTE}
            \mathcal{P}_{\tau,\text{MMV}_{\text{TE}}}[\bm{J}]:=\left\{\underset{\bm{S}}{\arg \min}\ \left\|\bm{J} - \bm{S}\right\|_F \ \text{s.t.}\ \kappa_{\text{TE}}(\bm{S}) \leq \tau \right\}.     
        \end{equation} 
        With a simple matrix transformation of $\overline{\bm{J}}_{n,:} = \left[\bm{J}_{2n-1,:}\ \bm{J}_{2n,:}\right]$ and $\overline{\bm{X}}_{n,:} = \left[\bm{X}_{2n-1,:}\ \bm{X}_{2n,:}\right]$, we can rewrite \eqref{eq.projectionTE} as follows
        \begin{equation}\label{eq.projectionTEtoTM}
            \left\{\underset{\overline{\bm{X}}}{\arg \min}\ \left\|\overline{\bm{J}}-\overline{\bm{X}}\right\|_F \ \text{s.t.}\ \left\|\overline{\bm{X}}\right\|_{1,2} \leq \tau \right\}=\mathcal{P}_{\tau,\text{MMV}_{\text{TM}}}\left[\overline{\bm{J}}\right].
        \end{equation}
        In doing so, the projection operator in TE case satisfies \cite[Theorem 6.3]{van2011sparse}. The (MMV$_{\text{TE}}$ BP$_{\widetilde\sigma}$) problem is solved by Algorithm \ref{alg.SPGTM} and Algorithm \ref{alg.NewtonRootTM} with the Pareto curve, its derivative with respect to $\tau$, and the projection operator replaced by $\phi_{\text{MMV}_{\text{TE}}}(\tau)$, $\phi'_{\text{MMV}_{\text{TE}}}(\tau)$, and $\mathcal{P}_{\tau,\text{MMV}_{\text{TE}}}[\cdot]$, respectively.    
        
    \subsection{CV-based Modified SPGL1}\label{subsec.CVSPGL1}

        In real applications, the noise level, i.e., the parameter $\widetilde\sigma$, is generally unknown, which means the termination condition, $\phi_{\text{MMV}}(\tau)=\widetilde\sigma$, does not work anymore. In order to deal with this problem, we modified the SPGL1 method based on the CV technique \cite{ward2009compressed,zhang2016cross}, in which $\widetilde\sigma$ is set 0 and the iteration is terminated using CV technique. In doing so, the problem of estimating the noise level, i.e., the parameter $\widetilde\sigma$, can be well circumvented. 

        Cross validation is a statistical technique that separates a data-set into a training (estimation) set and a testing (cross validation) set. The training set is used to construct the model and the testing set is used to adjust the model order so that the noise is not over-fitted. The basic idea behind this technique is to sacrifice a small number of measurements in exchange of prior knowledge. Specifically, when CV is utilized in SPGL1 method, we separate the original scattering matrix to a reconstruction matrix $\bm{\Phi}_{p,r}\in\mathbb{C}^{Q_r\times N}$ and a CV matrix $\bm{\Phi}_{p,CV}\in\mathbb{C}^{Q_{CV}\times N}$ with $Q = Q_r+Q_{CV}$. The measurement vector $\bm{y}_{p}$ is also separated accordingly, to a reconstruction measurement vector $\bm{y}_{p,r}\in\mathbb{C}^{Q_r}$ and a CV measurement vector $\bm{y}_{p,CV}\in\mathbb{C}^{Q_{CV}}$. The reconstruction residual and the CV residual are defined as
        \begin{equation}
            r_{\text{rec}} = \left(\sum_{p=1}^{P}\left\|\bm{y}_{p,r}-\bm{\Phi}_{p,r}\bm{j}_{p}\right\|_2^2\right)^{1/2}
        \end{equation}
        and
        \begin{equation}
            r_{\text{CV}} = \left(\sum_{p=1}^{P}\left\|\bm{y}_{p,CV}-\bm{\Phi}_{p,CV}\bm{j}_{p}\right\|_2^2\right)^{1/2},
        \end{equation}
        respectively. In doing so, every iteration can be viewed as two separate parts: reconstructing the contrast sources by SPGL1 and evaluating the outcome by the CV technique. The trend of CV residual in iteration behaves abruptly different (turns from decreasing to increasing) comparing to that of reconstruction residual, as soon as the reconstructed signal starts to overfit the noise. The reconstructed contrast sources are selected as the output on the criterion that its CV residual is the smallest one. 

        In order to find the smallest CV residual, a maximum number, $N_{\max}$, is needed and set a large value to guarantee the smallest CV residual occurs in the range of the $N_{\max}$ iterations. In this case, a large number of iterations are performed in vain, which decreases the efficiency of the algorithm. Therefore, we consider an alternative termination condition given by 
        \begin{equation}\label{eq.TerCond}
            N_{\text{Iter}} > N_{\text{opt}} + \Delta N,
        \end{equation}
        where, $N_{\text{Iter}}$ is the current iteration number, $N_{\text{opt}}$ is the iteration index corresponding to the smallest CV residual --- the optimal solution. The idea behind this criterion is that the CV residual is identified as the smallest one if the CV residual keeps increasing monotonously for $\Delta N$ times of iteration. In the following experimental examples, this termination condition works well with $\Delta N = 30$.

        Once the normalized contrast sources are obtained, one can achieve the shape of the scatterers defined as
        \begin{equation}\label{eq.MMVimageTM}
            \bm{\gamma}_{\text{MMV}_{\text{TM}}}[n] = \sum_{p=1}^P\left|\bm{j}_{p,n}\right|^2,
        \end{equation}
        or
        \begin{equation}\label{eq.MMVimageTE}
            \bm{\gamma}_{\text{MMV}_{\text{TE}}}[n] = \sum_{p=1}^P\left(\left|\bm{j}_{p,2n-1}\right|^2+\left|\bm{j}_{p,2n}\right|^2\right),
        \end{equation}
        with $n=1,2,\dots,N$, where $\bm{j}_{p,n}$, $\bm{\gamma}_{\text{MMV}_{\text{TM}}}[n]$, and $\bm{\gamma}_{\text{MMV}_{\text{TE}}}[n]$ represent the $n$-th element of vector $\bm{j}_{p}$, $\bm{\gamma}_{\text{MMV}_{\text{TM}}}$, and $\bm{\gamma}_{\text{MMV}_{\text{TE}}}$, respectively. In the end of this section, we remark that as the regularized solution corresponds to the least sum-of-norm, the non-measurable equivalent contrast sources \cite{caorsi1999inverse} tend to be ignored.

\section{Linear sampling method and its improved version}\label{sec.LSM}

    In this section, the proposed method is tested with both synthetic data and experimental data. In the meanwhile, we have also processed the same data using linear sampling method for comparison. Since the background of the experiments is free space, the LSM method consists in solving the integral equation of the indicator function
    \begin{equation}\label{eq.LSM.TM}
        \int E_3( \bm{x}_r, \bm{x}_t)g_3( \bm{x}_s, \bm{x}_t)d \bm{x}_t = E_{3,3}( \bm{x}_s, \bm{x}_r),
    \end{equation}
    and
    \begin{equation}\label{eq.LSM.TE}
       \int 
        \begin{bmatrix}
            E_1 & 0\\
            0 & E_2
        \end{bmatrix}( \bm{x}_r, \bm{x}_t)
        \begin{bmatrix}
            g_{1,1} & g_{1,2}\\
            g_{2,1} & g_{2,2}
        \end{bmatrix}( \bm{x}_s, \bm{x}_t)d \bm{x}_t 
        = 
        \begin{bmatrix}
            E_{1,1} & E_{1,2}\\
            E_{2,1} & E_{2,2}
        \end{bmatrix}( \bm{x}_s, \bm{x}_r)
    \end{equation}
    for the TM and TE cases, respectively, where, $E_1(\bm{x}_r, \bm{x}_t)$, $E_2(\bm{x}_r, \bm{x}_t)$, and $E_3(\bm{x}_r, \bm{x}_t)$ represent \textcolor{red}{$x_1$-, $x_2$-, and $x_3$}-components of the scattered field probed at $\bm{x}_r$ corresponding to the transmitter at $ \bm{x}_t$, respectively; $ \bm{x}_s$ is the sampling point in the inversion domain; $E_{i,j}( \bm{x}_s, \bm{x}_r)$ is $i$-th component of the electric field at $\bm{x}_r$ generated by an ideal electric dipole located at $\bm{x}_s$ with the polarization vector parallel to $x_j$-axis, which are given by
    \begin{subequations}
      \begin{align}
          E_{3,3}&= \frac{1}{4}\omega\mu_0H_0^{(1)}(-kR),\\
          E_{1,1}&= \frac{-k}{4\omega\varepsilon_0}\left(\frac{H_1^{(1)}(-kR)}{R}+\frac{kx_2^2}{R^2}H_2^{(1)}(-kR)\right),\\
          E_{1,2}&= \frac{k^2x_1x_2}{4\omega\varepsilon_0R^2}H_2^{(1)}(-kR),\\
          E_{2,1}&= \frac{k^2x_1x_2}{4\omega\varepsilon_0R^2}H_2^{(1)}(-kR),\\
          E_{2,2}&= \frac{-k}{4\omega\varepsilon_0}\left(\frac{H_1^{(1)}(-kR)}{R}+\frac{kx_1^2}{R^2}H_2^{(1)}(-kR)\right),
      \end{align}
    \end{subequations} 
    where, $R=\| \bm{x}_s- \bm{x}_r\|_2$. Eq.~\eqref{eq.LSM.TM} and \eqref{eq.LSM.TE} can be reformulated as a set of linear systems of equations
    \begin{equation}\label{eq.LSMEq}
        \bm{Y}\bm{g}_{ \bm{x}_s}=\bm{f}_{ \bm{x}_s},
    \end{equation}
    where, $\bm{Y}$ is the measurement data matrix, $\bm{g}_{ \bm{x}_s}$ is the indicator function of the sampling point $ \bm{x}_s$ in the form of a column vector, $\bm{f}_{ \bm{x}_s}$ is the right side of Eq.~\eqref{eq.LSM.TM} in the form of a column vector. Following the same approach of solving Eq.~\eqref{eq.LSMEq} in \cite{catapano2007simple,crocco2012linear}, the shape of the scatterers is defined by 
    \begin{equation}\label{eq.LSMimage}
        \bm{\gamma}_{\text{LSM}}( \bm{x}_s) = \frac{1}{\|\bm{g}_{ \bm{x}_s}\|^2},
    \end{equation}
    where, $\|\bm{g}_{ \bm{x}_s}\|^2$ is given by
    \begin{equation}
        \|\bm{g}_{ \bm{x}_s}\|^2 = \sum_{d=1}^D\left(\frac{s_{d}}{s_{d}^2+a^2}\right)^2\left|\bm{u}_d^H\bm{f}_{ \bm{x}_s}\right|^2,
    \end{equation}
    where, $s_{d}$ represents the singular value of matrix $\bm{Y}$ corresponding to the singular vector $\bm{u}_d$, $D=\min\{P,Q\}$, and $a=0.01\times\underset{d}{\max}\{s_{d}\}$. 

    We have also considered, in the TM cases, the improved linear sampling method proposed in \cite{crocco2013improved} in the comparison of the proposed method and LSM. The indicator function of improved LSM is defined as
    \begin{equation}
      \bm{\gamma}_{\text{LSM},I}( \bm{x}_s) = \left(\prod_{i=1}^{I}\frac{\|\bm{g}^{x}_{i, \bm{x}_s}\|^2}{\|\bm{g}_{ \bm{x}_s}\|^2}\frac{\|\bm{g}^{y}_{i, \bm{x}_s}\|^2}{\|\bm{g}_{ \bm{x}_s}\|^2}\right)^{\frac{1}{2I}},\quad I = ka,
    \end{equation}
    where, $a$ is the radius of a smallest ball that covers the targets, the power $\frac{1}{2I}$ is normalization factor, and $\bm{g}^{x}_{i, \bm{x}_s}$ and $\bm{g}^{y}_{i, \bm{x}_s}$ are obtained by replacing $E_{3,3}( \bm{x}_s, \bm{x}_r)$ in Eq. \eqref{eq.LSM.TM} with $\varphi^x_i(\bm{x}_s, \bm{x}_r)$ and $\varphi^y_i( \bm{x}_s, \bm{x}_r)$, respectively,
    \begin{subequations}
      \begin{align}
        \varphi^x_i&=\frac{1}{4}\omega\mu_0H_i^{(1)}(-kR)\cos\left(i(\phi_r-\phi_s)\right),\\
        \varphi^y_i&=\frac{1}{4}\omega\mu_0H_i^{(1)}(-kR)\sin\left(i(\phi_r-\phi_s)\right).
      \end{align}
    \end{subequations}
    where, $\phi_r$ and $\phi_s$ are the angular components of $ \bm{x}_s$ and $ \bm{x}_r$ in polar coordinate system, respectively. We refer to \cite{crocco2013improved} for more details of this indicator function.

    It is worth mentioning that both the contrast source $\bm{j}_{p}$ and the indicator function $\bm{g}_{ \bm{x}}$ are proportional to the amplitude of the electric field. According to the definition in Eq.~\eqref{eq.MMVimageTM}, Eq. \eqref{eq.MMVimageTE}, and Eq.~\eqref{eq.LSMimage}, $\bm{\gamma}_{\text{MMV}}$ and $\bm{\gamma}_{\text{LSM}}$ are proportional and inversely proportional to the power of the electric field, respectively. Therefore, the dB scaling shown in the following examples is defined as
    \begin{equation}
        \bm{\gamma}_{\text{dB}}=10\times \log_{10}\left(\frac{\bm{\gamma}}{\max\{\bm{\gamma}\}}\right).
    \end{equation}

\section{Synthetic Data Imaging}\label{sec.Syn}

    \begin{figure}[!t]
        \centering
        \includegraphics[height=0.45\linewidth]{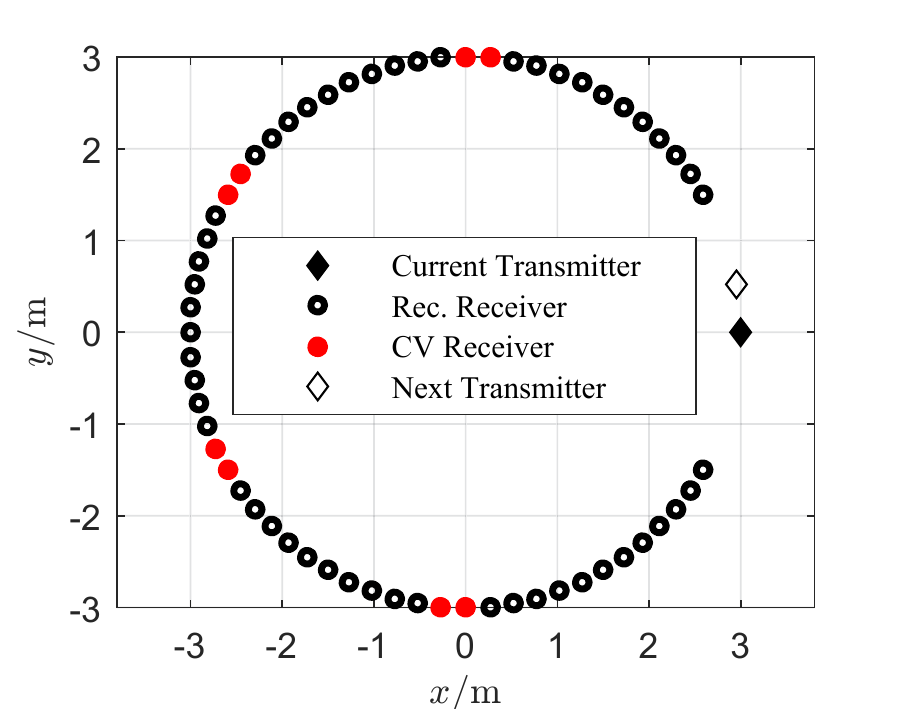}
        \caption{Measurement configuration of Simulation 1 and 2.}
        \label{fig:MMV_Sim_Conf}
    \end{figure}

    In this subsection, the proposed method is tested with synthetic data. The transmitting antenna is simulated for simplicity with an ideal electric dipole (TM-polarization case) and an ideal magnetic dipole (TE-polarization case). Coordinate system is established such that the dielectric parameters are variable along the $x$- and $y$-axes, but invariable along $z$-axis. The transmitting antenna rotates on a circular orbit of 3 m radius centering at the origin (0,0). The receiving positions are taken on the same orbit without any position close than 30$^\circ$ from the transmitting antenna. The measurement configuration of Simulation 1 and 2 is shown in Figure \ref{fig:MMV_Sim_Conf}, in which the selection of CV measurements and reconstruction measurements is illustrated. Empirically, an arc length $\geq \lambda/3$ is a good selection. The number of the CV receivers on each arc depends on how dense the receiver positions are, and the total CV receiver number is around 20\% of the total measurement number \cite{zhang2016cross}. The operating frequency is 500 MHz. Two configurations of different objects are considered. One is combined with two circular metallic cylinders and the other one is a metallic cylinder with a ``crescent-shaped'' cross section. The radius of the circular cross section is $0.2$ m ($=\lambda/3$), and the centers of the two circles are ($-0.45$, $0.6$) and ($0.45$, $0.6$), respectively. The crescent is the subtraction of two circles of radius $0.6$ m ($=\lambda$) centering at ($0$, $0$) and ($0.4$, $0$). See Figure~\ref{fig:MMVSim1}(a) and Figure~\ref{fig:MMVSim2}(a) for their true geometry. The forward EM scattering problem is solved by a MATLAB-based 3-D FDFD package ``MaxwellFDFD'' \cite{W.Shin2013}. The technique of non-uniform staggered grids is used to reduce the computational burden, while for inverting the measurement data, we consider uniform discretization such that an inverse crime is circumvented. In the forward solver, we consider a fine grid size of $\lambda/(45\sqrt{\epsilon_r})$. The data for inversion is obtained by subtracting the incident field from the total field. Periodic boundary conditions are imposed on the design of the FDFD stiffness matrix in order to simulate the 2-D configuration. Perfect matching layer (PML) is used to simulate an anechoic chamber environment. 

    \subsection{Determine the measurement configuration}\label{subsec.det.mea.conf}

        \begin{figure}[!t]
            \centering
            \subfloat[]{\includegraphics[width=0.50\linewidth] {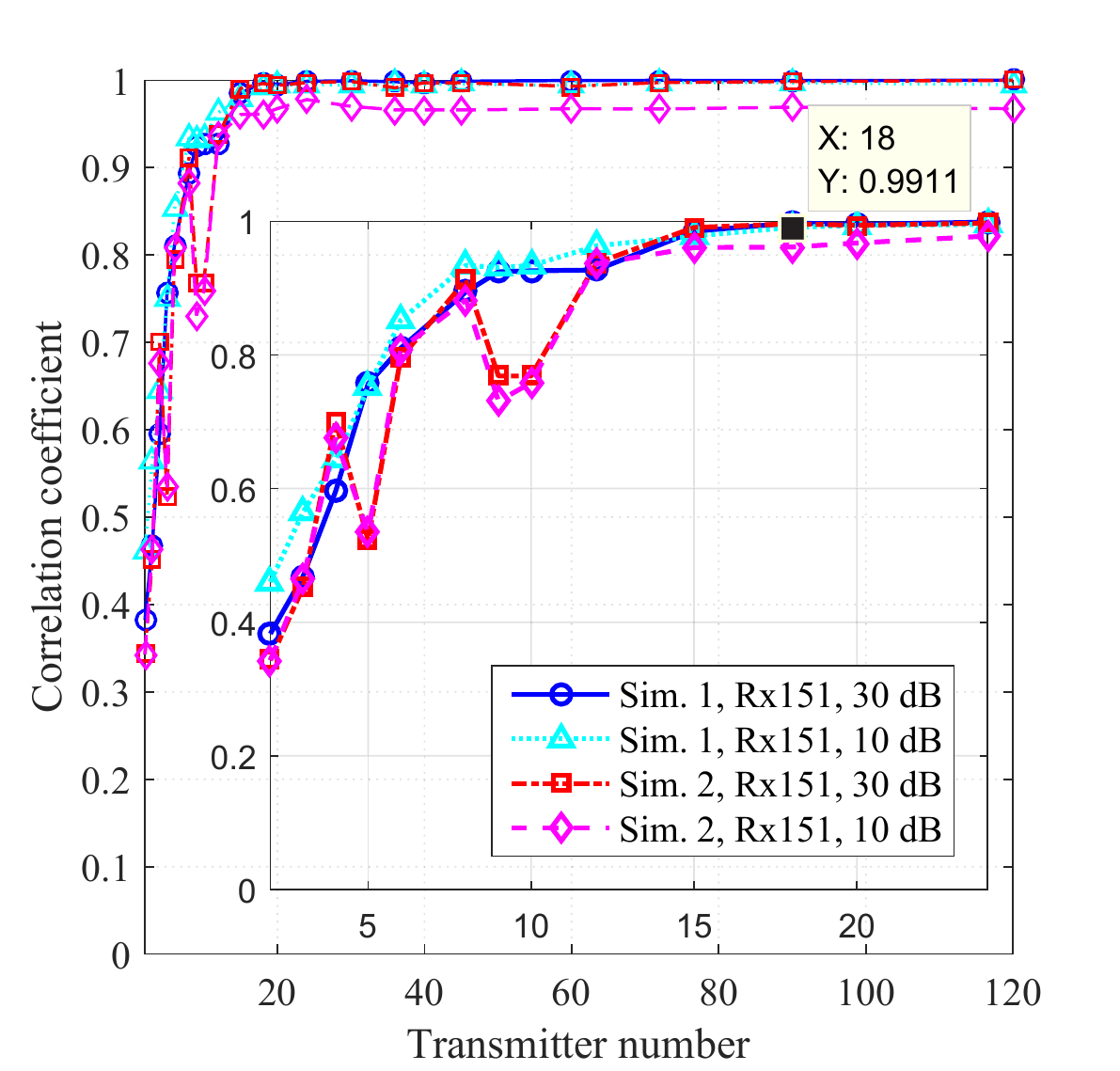}}%
            \subfloat[]{\includegraphics[width=0.50\linewidth] {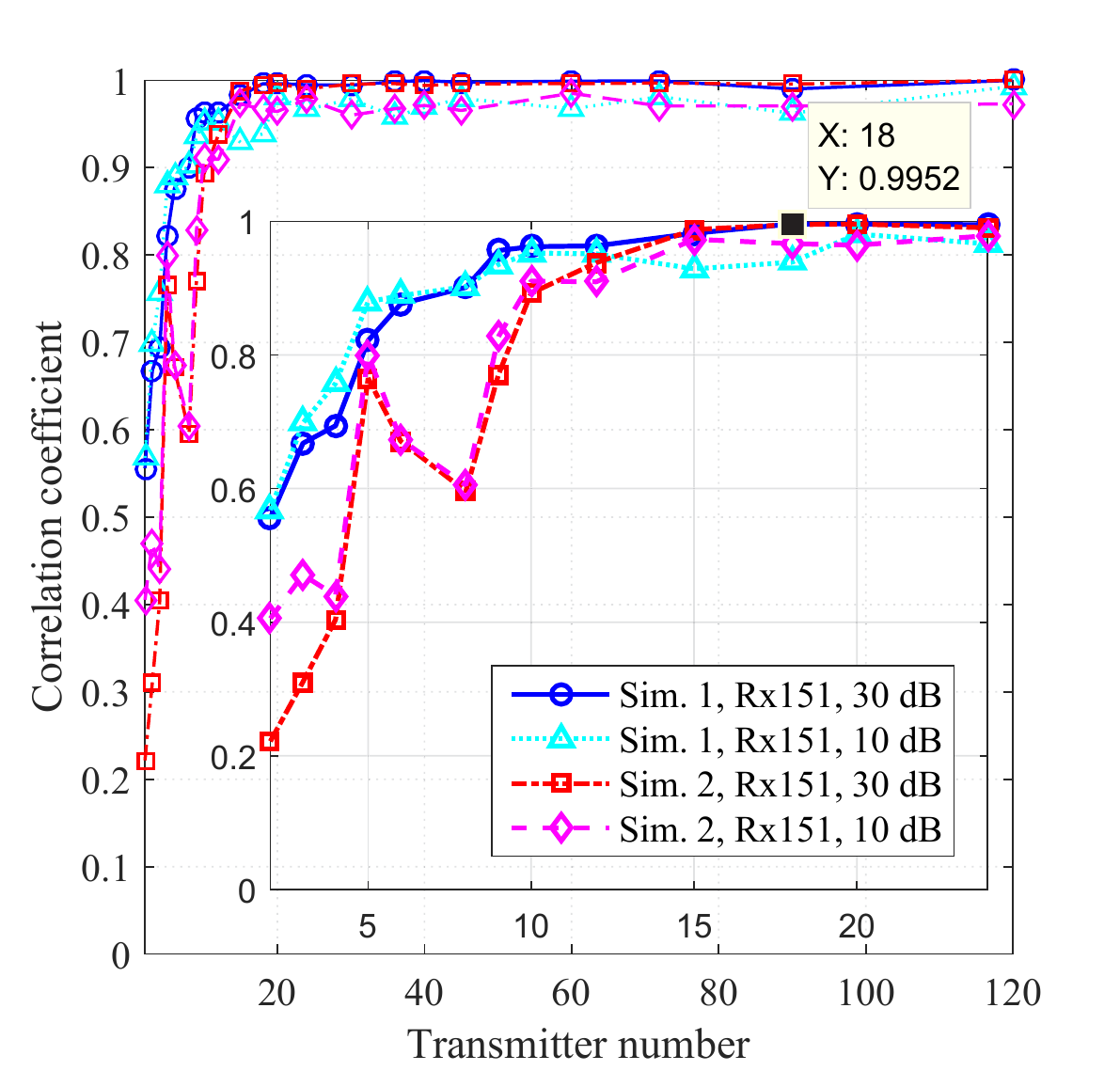}}
            \caption{Correlation coefficient curves in terms of transmitter number in Simulation 1 and 2. Receiver number is fixed to 151. 10dB and 30 dB Gaussian random additive noises are considered, respectively. (a) TM-polarized data; (b) TE-polarized data.}
            \label{fig:MatchTx_Tx}
        \end{figure}

        \begin{figure}[!t]
            \centering
            \subfloat[]{\includegraphics[width=0.50\linewidth] {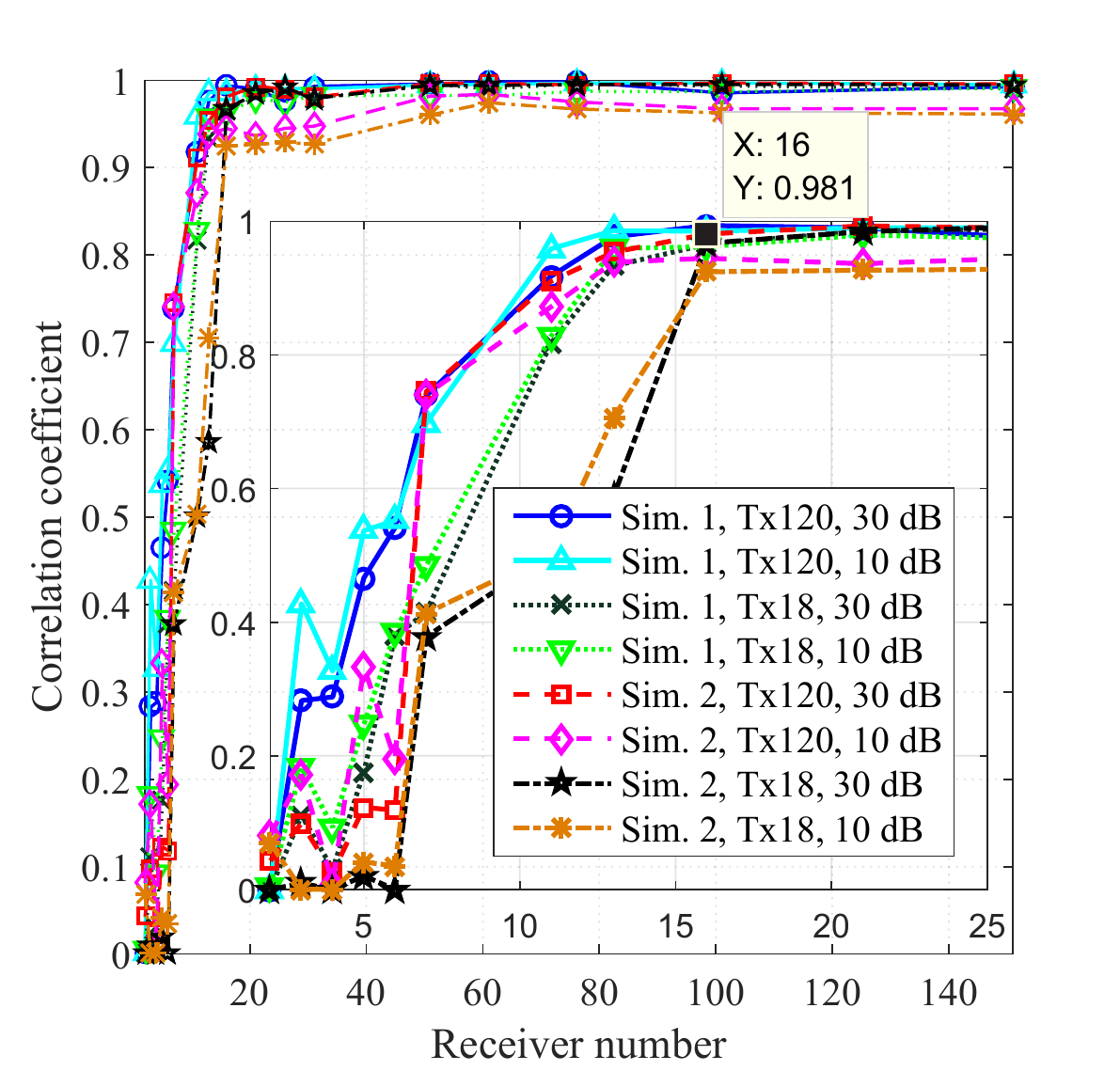}}%
            \subfloat[]{\includegraphics[width=0.50\linewidth] {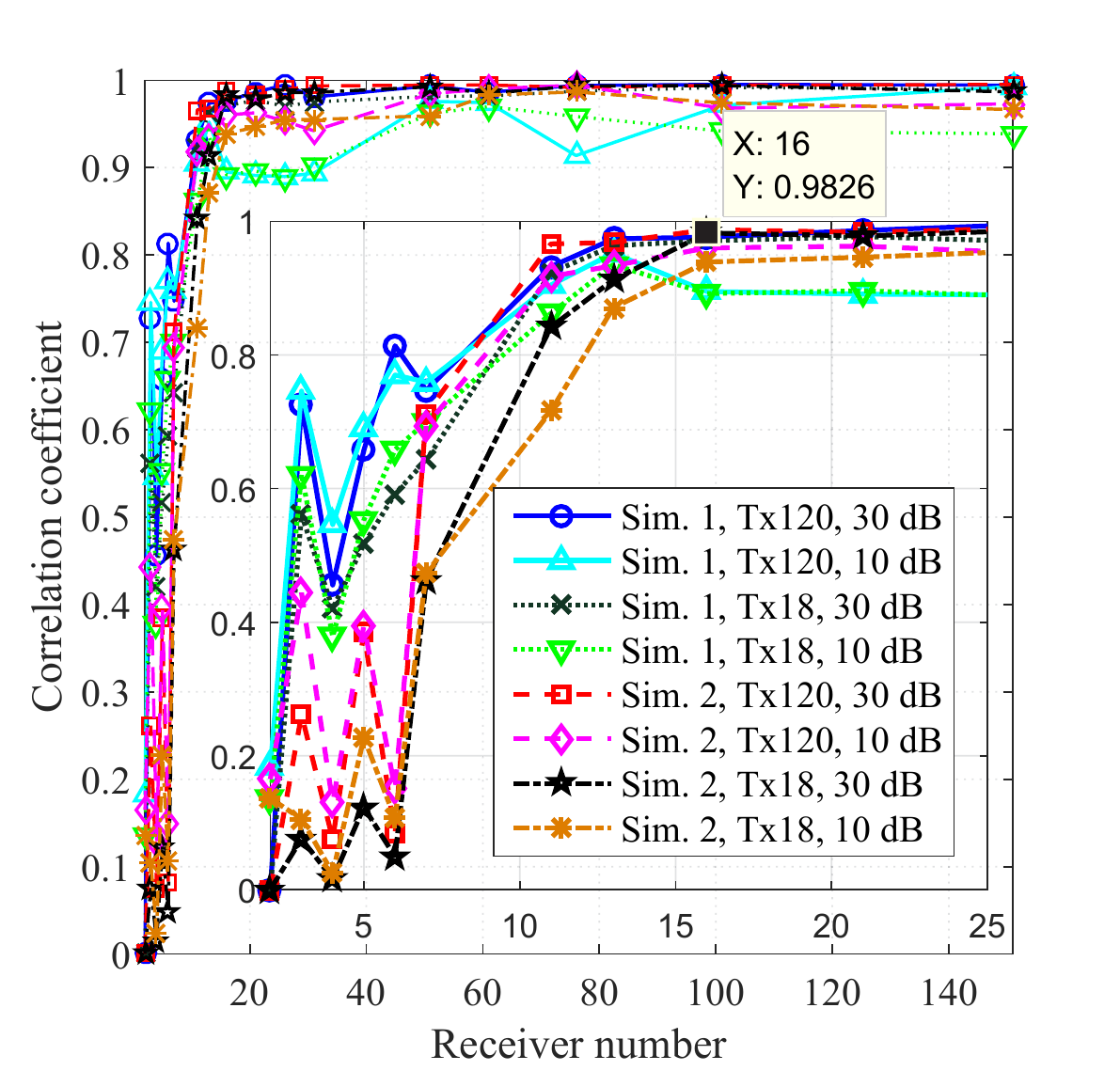}}
            \caption{Correlation coefficient curves in terms of receiver number in Simulation 1 and 2. Transmitter number is fixed to 18 and 120, respectively. 10dB and 30 dB Gaussian random additive noises are considered, respectively. (a) TM-polarized data; (b) TE-polarized data.}
            \label{fig:MatchTx_Rx}
        \end{figure}

        To determine the measurement configuration, we need to investigate the relationship between the transceiver numbers and the imaging quality. Let us first consider 120 transmitters and 151 receivers, i.e., the transmitter rotates on the circular orbit with a step of 3$^\circ$, and the receiver rotates on the measurement arc of 300$^\circ$ with a step of 2$^\circ$. The CV receivers are selected in the same way shown in Figure \ref{fig:MMV_Sim_Conf}, but there are 4 continuous CV receiver positions in each CV arc (equivalent to 8 $^\circ$). Now let us disturb the measurement data (the scattered fields) with Gaussian additive random noise of 30 dB signal-to-noise ratio (SNR), and then process the data by the proposed method. If we use the reconstructed image as the reference image, denoted by $\bm{\gamma}_\text{ref}$, then a correlation coefficient can be defined as
        \begin{equation}
          r_{\text{corr}}:=\frac{\sum_{n=1}^N\left(\bm{\gamma}_\text{ref}[n]-\overline{\gamma_\text{ref}}\right)\left(\bm{\gamma}[n]-\overline{\gamma}\right)}{\sqrt{\sum_{n=1}^N\left(\bm{\gamma}_\text{ref}[n]-\overline{\gamma_\text{ref}}\right)^2 \sum_{n=1}^N\left(\bm{\gamma}[n]-\overline{\gamma}\right)^2}},
        \end{equation}         
        where, $\bm{\gamma}$ denotes the MMV image with different measurement configurations and noise levels, $\overline{\gamma_\text{ref}}$ and $\overline{\gamma}$ are the mean values of $\bm{\gamma}_\text{ref}$ and $\bm{\gamma}$, respectively. The correlation coefficient reflects the similarity degree of two images. The minor negative correlation coefficients are forced to zeros, as negative correlation does not make any sense for two amplitude images.

        Now we first fix the receiver number to 151, and calculate the correlation coefficients of Simulation 1 and Simulation 2 with different transmitter numbers. Fig.~\ref{fig:MatchTx_Tx} (a) and (b) show the correlation coefficient curves in terms of transmitter number by processing the TM-polarized data and TE-polarized data, respectively. Two SNRs, 10 dB and 30 dB, are considered. From Fig.~\ref{fig:MatchTx_Tx} we observe that obvious decrease of correlation coefficient occurs at 18 transmitters, indicating that the image quality gets worse when transmitter number is less than 18. The correlation coefficient curves of 10 dB and 30 dB maintain the same trend, and the correlation coefficients of 10 dB maintains above 0.95 when more than 18 transmitters are used, indicating the proposed method is robust against the Gaussian additive random noise. Then we fix the transmitter number to 18 and 120, respectively, and image the targets in Simulation 1 and 2 with different receiver numbers. Since CV technique needs enough amount of measurements, the noise level is assumed exactly known when the receiver number is less than or equal to 31. Fig.~\ref{fig:MatchTx_Rx}(a) and (b) show the correlation coefficient curves in terms of receiver number by processing the TM-polarized data and TE-polarized data, respectively. Two SNRs, 10 dB and 30 dB, are considered. From Fig.~\ref{fig:MatchTx_Rx} we observe that the smallest receiver number for ensure a stable imaging quality is 16. The correlation coefficient of 18 transmitters and 10 dB SNR maintains $r_{\text{corr}}\geq 0.90$ when the receiver number $\geq 16$. 

        Since the reference image in the definition of the correlation coefficient is not the real shape of the targets, it is actually an asymptotic measure of the imaging quality. Discussion of the imaging results are given in the next subsection in comparison to the LSM images to further investigate the imaging performance. In the following subsection of imaging results and  the next section of experimental data imaging, we select 18 transmitters (equivalent to an interval of 20$^\circ$) for two reasons: 1) the proposed method works well in the numerical simulations with 18 transmitters; 2) more experiments and targets are required to demonstrate the good imaging performance when 18 transmitters are used. Note that the noise level is not available in real applications, 61 receivers (equivalent to an interval of 5$^\circ$) are selected in the numerical experiments for the use of CV technique.   

    \subsection{Imaging results}

    \subsubsection{Simulation 1}
        \begin{figure}[!t]
            \centering
            \makebox[\columnwidth]{
            \subfloat[]{\includegraphics[height=0.375\linewidth] {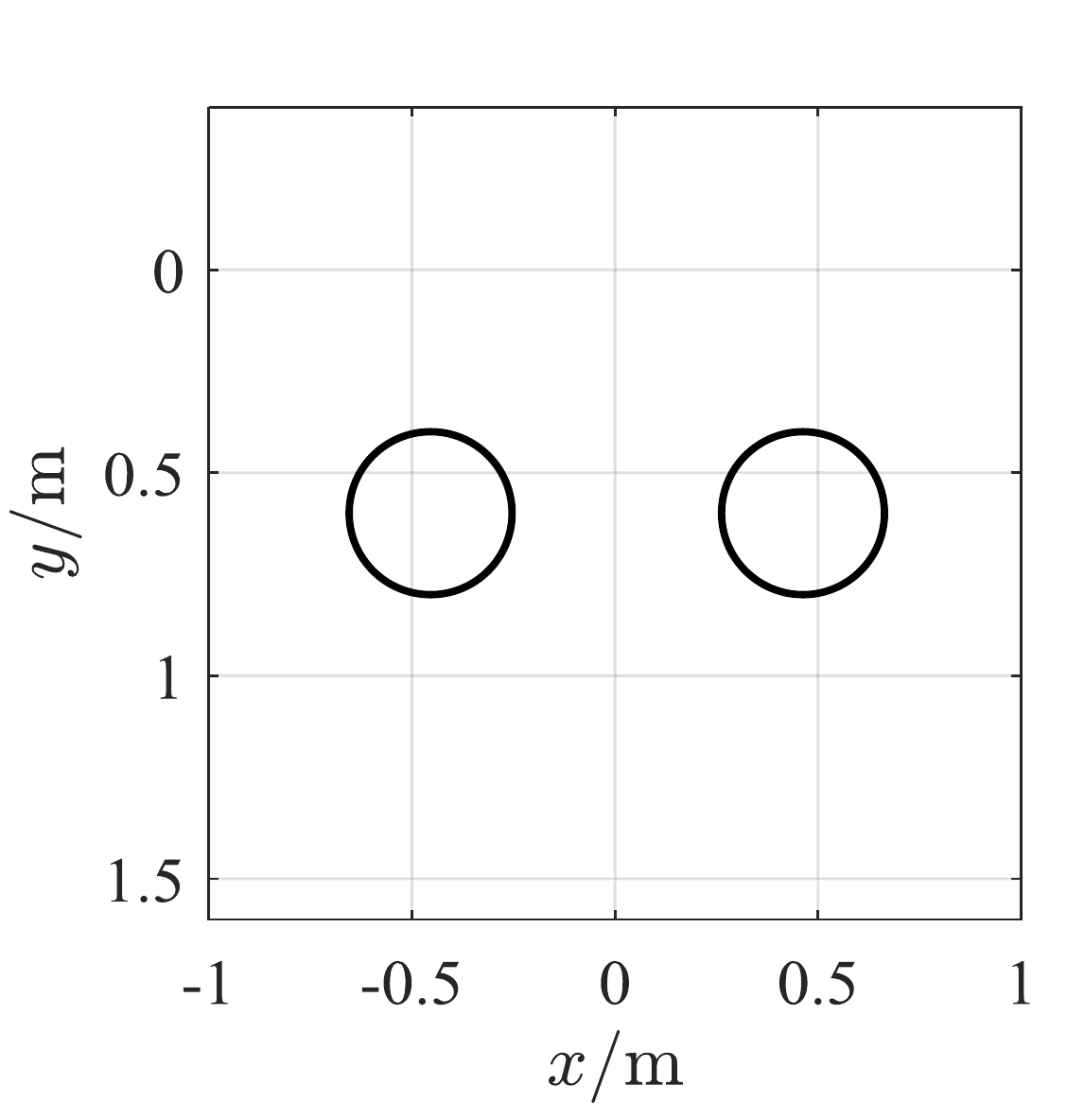}}\qquad
            \subfloat[]{\includegraphics[height=0.375\linewidth] {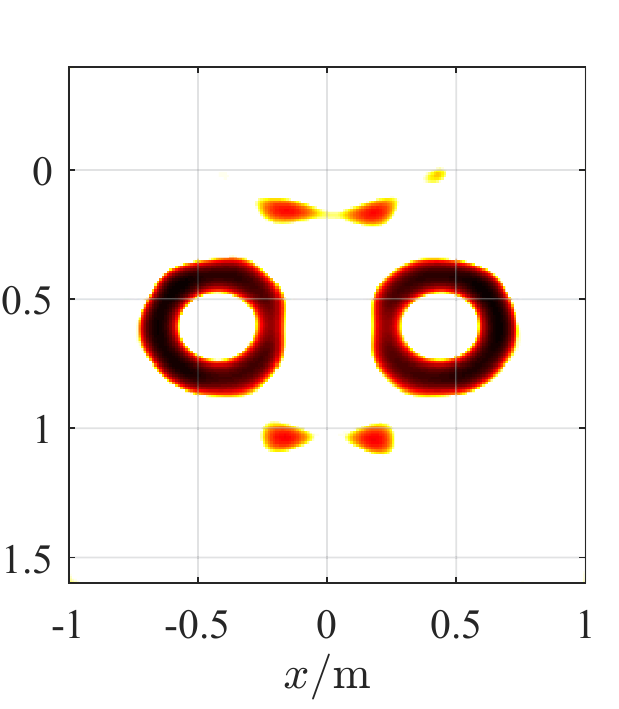}}} 
            \makebox[\columnwidth]{
            \subfloat[]{\includegraphics[height=0.375\linewidth] {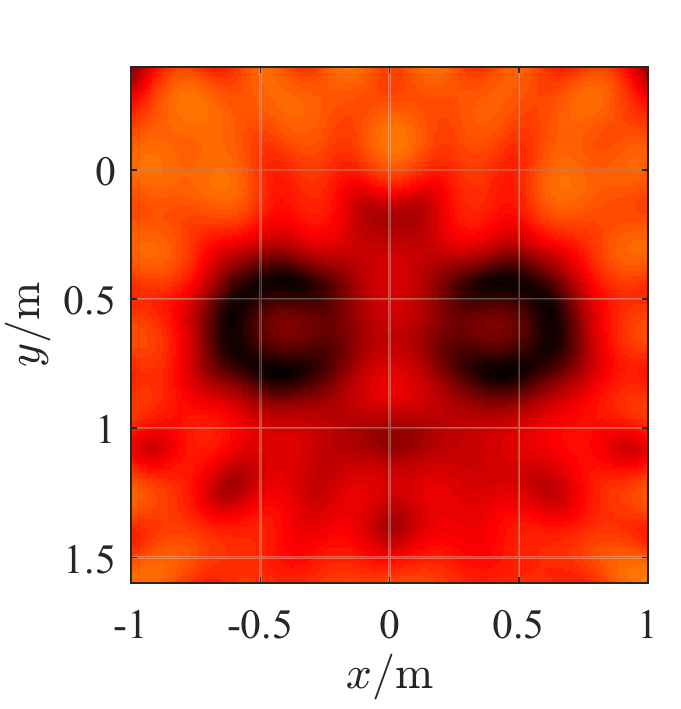}}\qquad
            \subfloat[]{\includegraphics[height=0.375\linewidth] {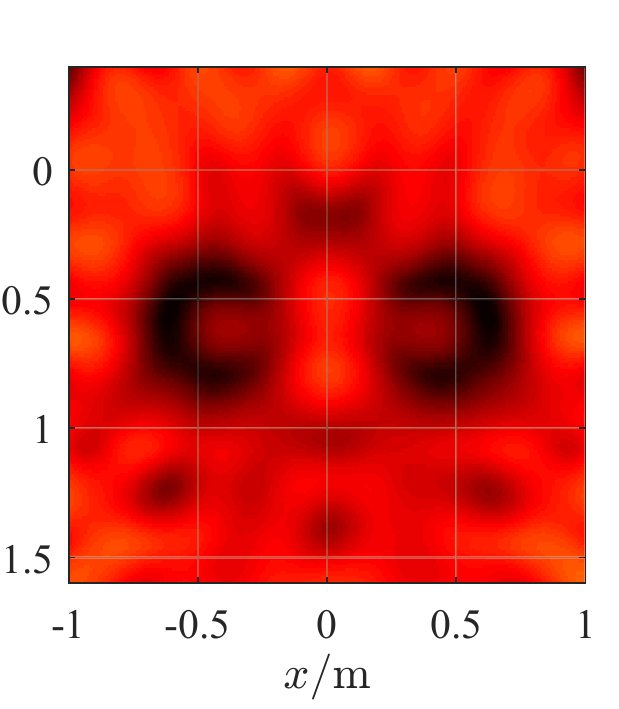}}}
            \raggedright
            \includegraphics[width=0.35\linewidth] {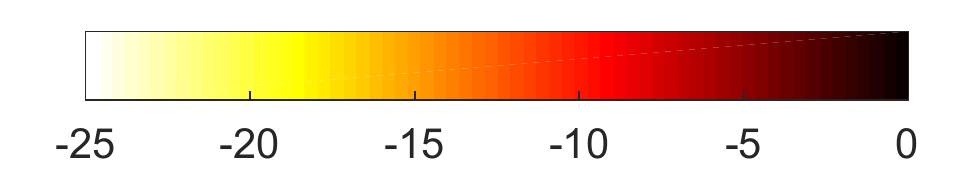}
            \caption{Scatterer geometry and its reconstructed shapes in Simulation 1. 30 dB Gaussian noise is added to the measurement data. (a) Scatterer geometry; The scatterer shape (the value of the indicator function in dB) reconstructed by processing the TM-polarized data with MMV (b), LSM (c), and improved LSM with $I=7$ (d), respectively.}
            \label{fig:MMVSim1}
        \end{figure}

        \begin{figure}[!t]
            \centering
            \makebox[\columnwidth]{
            \subfloat[]{\includegraphics[height=0.375\linewidth] {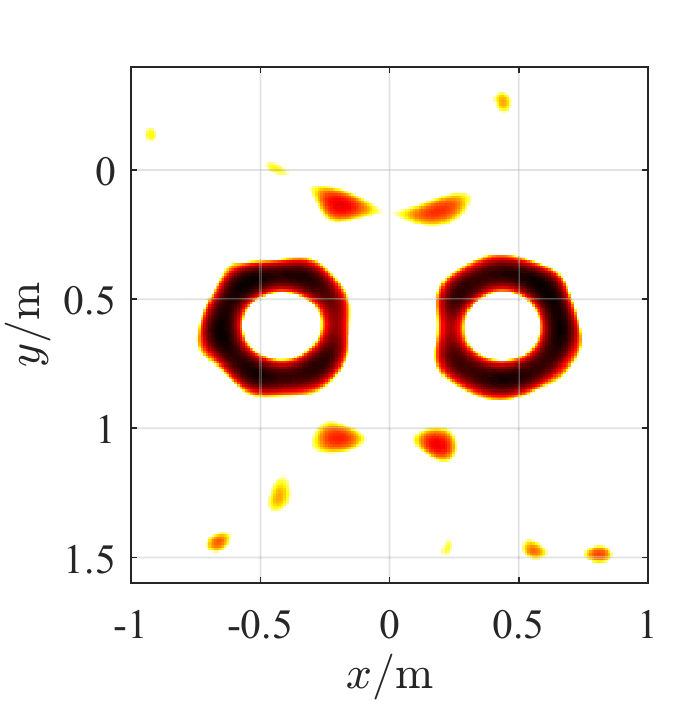}}} 
            \makebox[\columnwidth]{
            \subfloat[]{\includegraphics[height=0.375\linewidth] {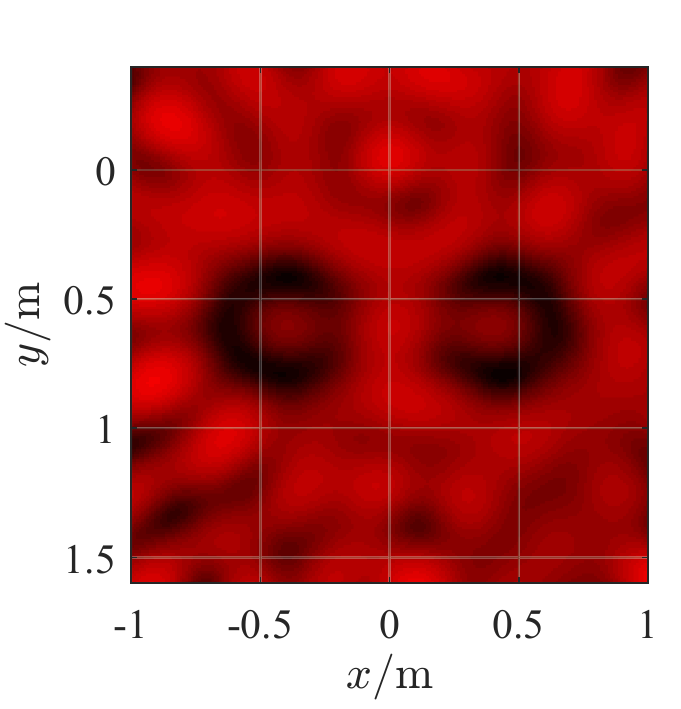}}\qquad
            \subfloat[]{\includegraphics[height=0.375\linewidth] {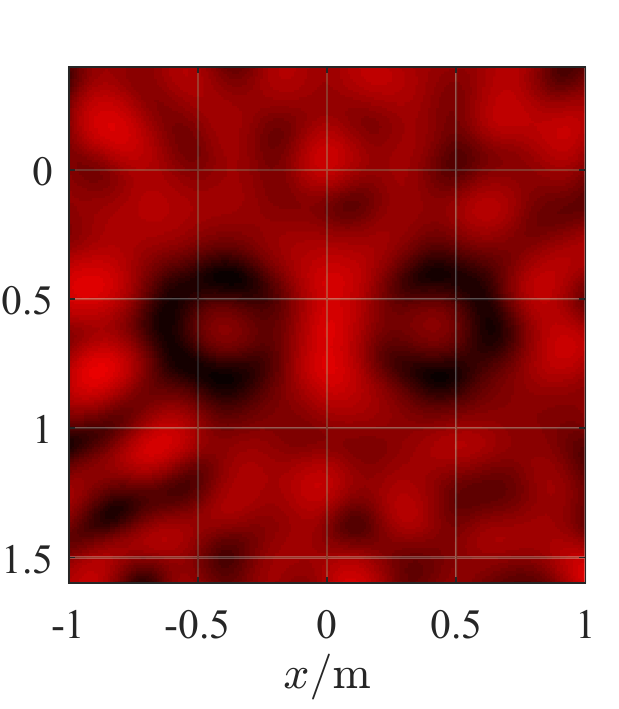}}}
            \raggedright
            \includegraphics[width=0.35\linewidth] {colorbar}
            \caption{Scatterer geometry and its reconstructed shapes in Simulation 1. 10 dB Gaussian noise is added to the measurement data. The scatterer shape (the value of the indicator function in dB) reconstructed by processing the TM-polarized data with MMV (a), LSM (b), and improved LSM with $I=7$ (c), respectively.}
            \label{fig:MMVSim1TM10dB}
        \end{figure}

        \begin{figure}[!t]
            \centering
            \subfloat[]{\includegraphics[width=0.50\linewidth] {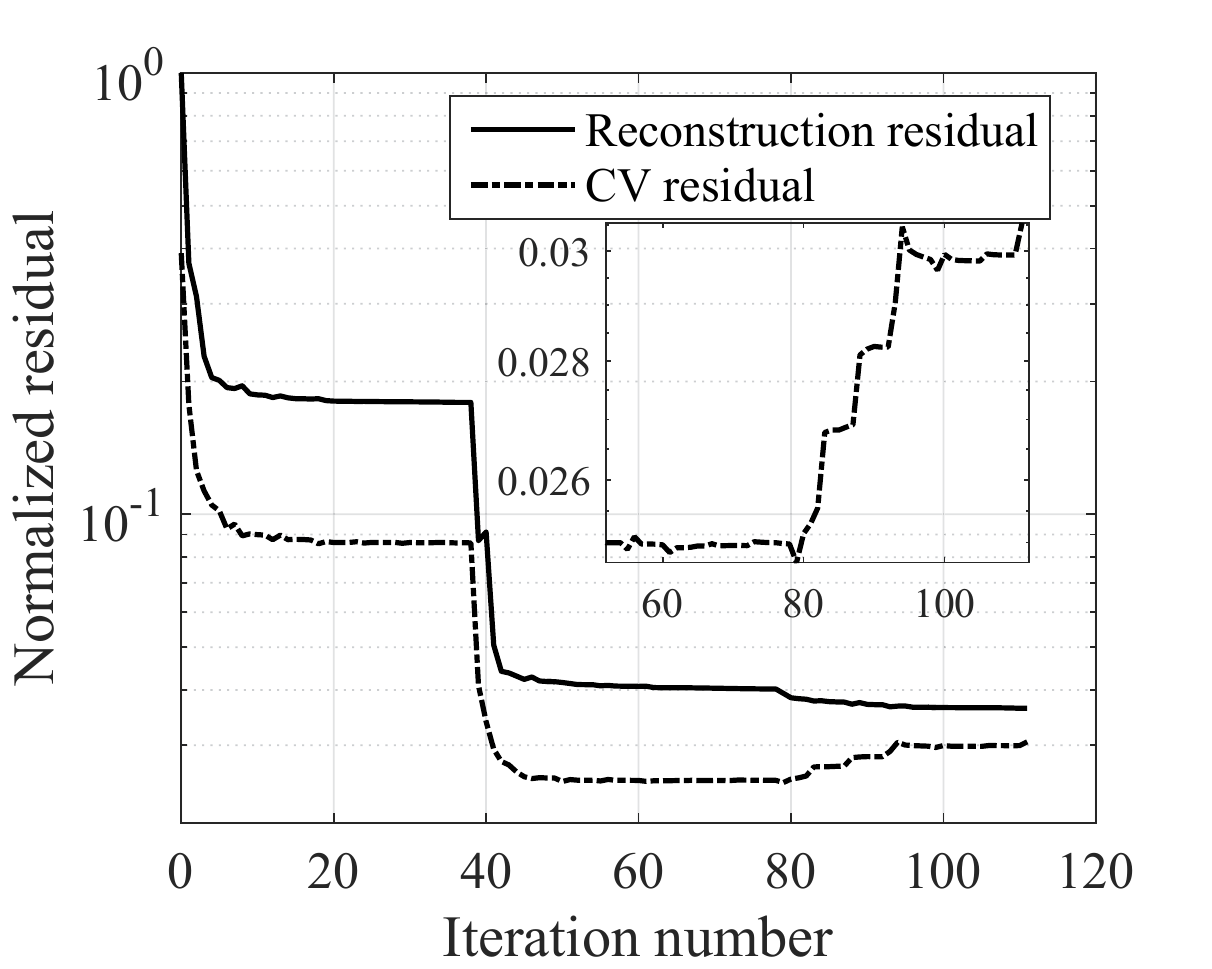}}%
            \subfloat[]{\includegraphics[width=0.50\linewidth] {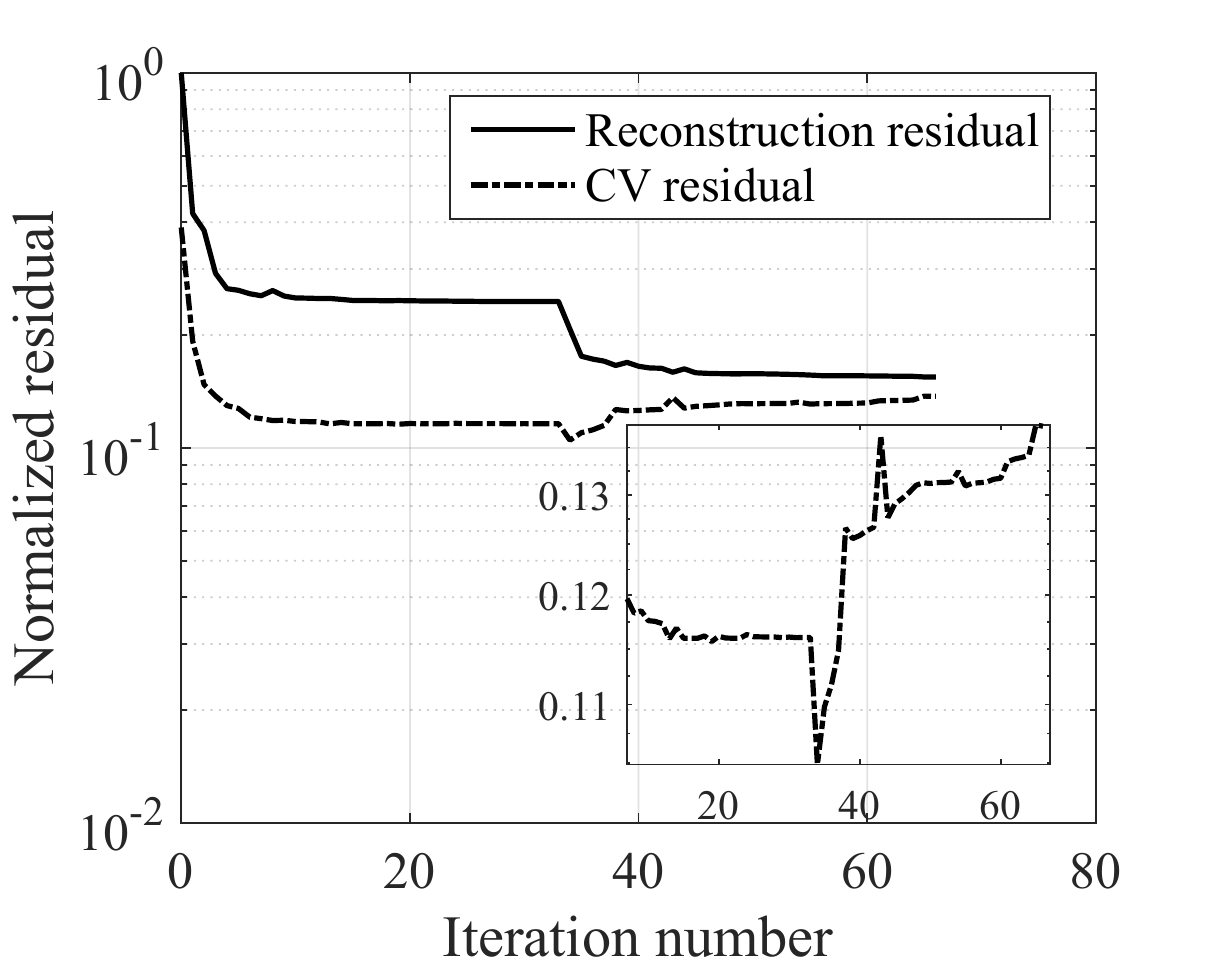}}
            \caption{Reconstruction residual and CV residual curves by processing the TM-polarized data of Simulation 1. (a) SNR$=30$ dB; (b) SNR$=10$ dB.}
            \label{fig:MMVSim1_CV}
        \end{figure}

        \begin{figure}[!t]
            \centering
            \makebox[\columnwidth]{
            \subfloat[]{\includegraphics[height=0.375\linewidth] {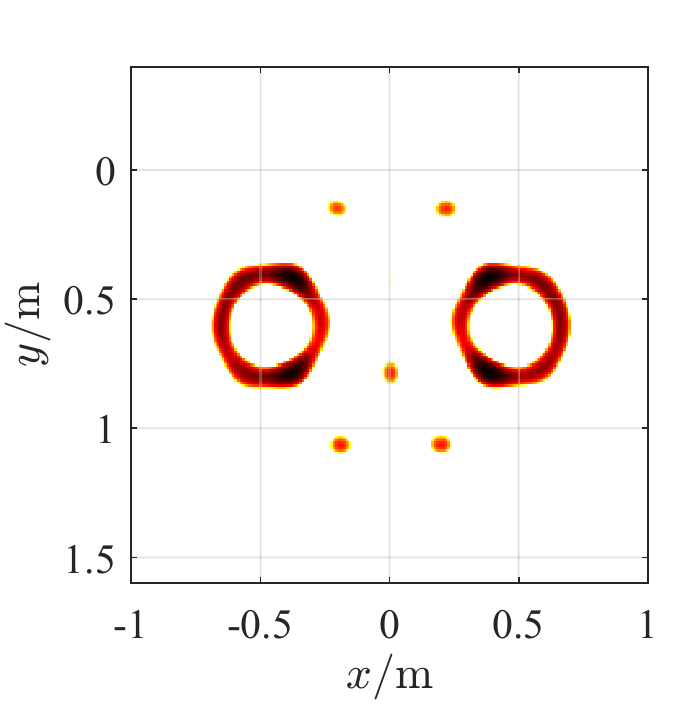}}\qquad
            \subfloat[]{\includegraphics[height=0.375\linewidth] {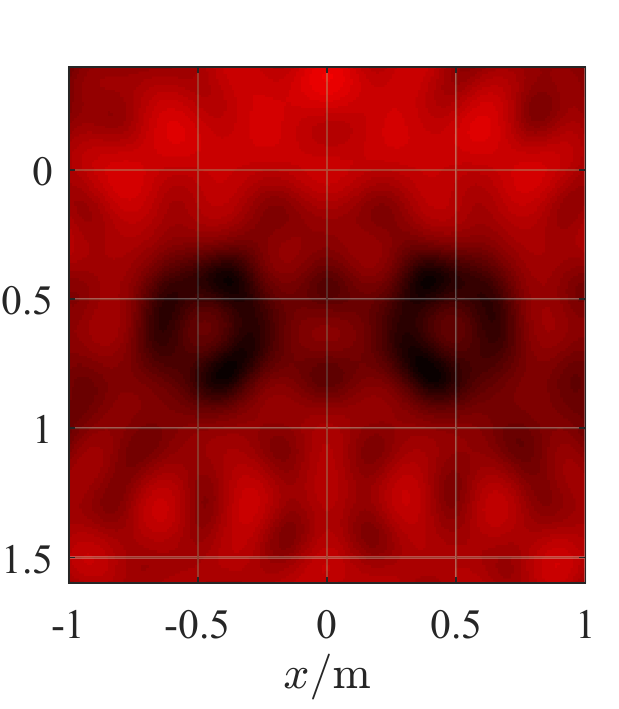}}}
            \makebox[\columnwidth]{
            \subfloat[]{\includegraphics[height=0.375\linewidth] {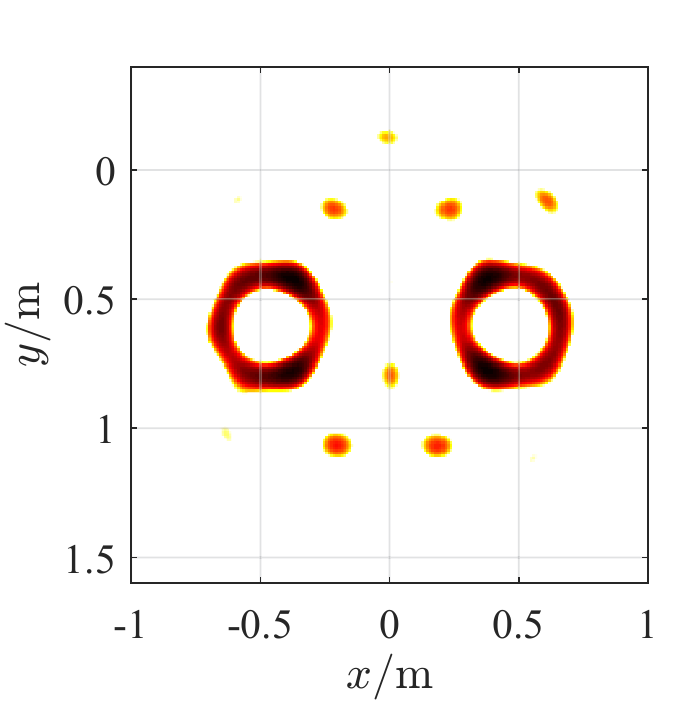}}\qquad
            \subfloat[]{\includegraphics[height=0.375\linewidth] {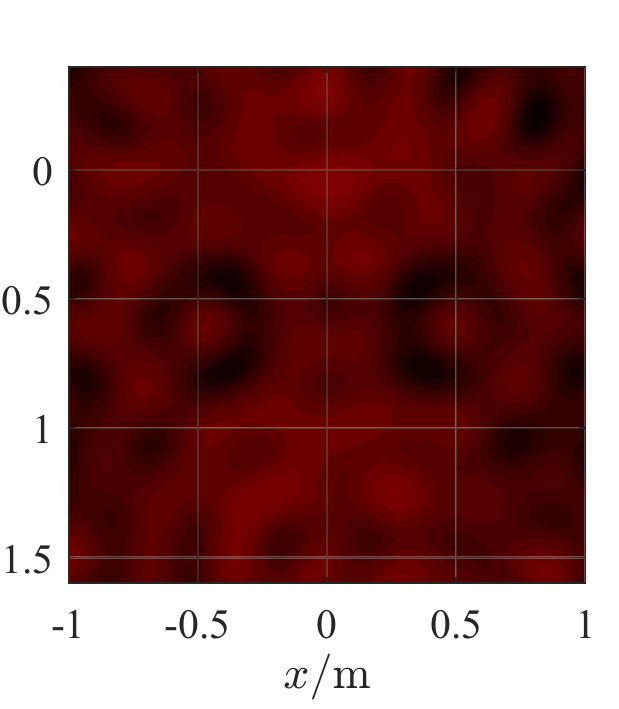}}}
            \raggedright
            \includegraphics[width=0.35\linewidth] {colorbar}
            \caption{Scatterer shape (the value of the indicator function in dB) reconstructed by processing the TE-polarized data of Simulation 1. (a), (c): MMV images; (b), (d): LSM images; (a), (b): SNR$=30$ dB; (c), (d): SNR$=10$ dB.}
            \label{fig:MMVSim1TE}
        \end{figure}

         To reduce the computational cost, we restrict the inversion domain to $[-1.0,1.0] \times [-0.4,1.6]$ m$^2$. The inversion domain is discretized with equal grid size of $0.01$ m ($=\lambda/60$). Let us first consider the TM-polarized data disturbed with Gaussian additive random noise of 30 dB SNR. The residual curves are shown in Figure \ref{fig:MMVSim1_CV}(a). The trend of the residual curves is like staircases, and each step corresponds to one update of the parameter $\tau$. The CV residual starts to increase after 80 iterations, and $\Delta N=30$ more iterations are performed before termination. The solution of the minimum CV residual is selected as the optimal solution. The scatterer shape reconstructed by MMV, LSM, and improved LSM with $I=7$ is shown in Figure \ref{fig:MMVSim1}(b), (c) and (d), respectively. By comparison of Figure \ref{fig:MMVSim1}(c) and (d), it is observed that the artifacts between the two circular cylinders are suppressed by improved LSM. However, the average amplitude of the sidelobes in the region of no targets increases from $-15$ dB to $-10$ dB. From Figure \ref{fig:MMVSim1}(b) we observe that the proposed method shows higher resolution and lower sidelobes in comparison to \ref{fig:MMVSim1}(c) and (d), indicating that the resolving ability of the proposed method is better than LSM. To study the imaging performance with different SNRs, let us decrease the SNR to 10 dB, and the images results are shown in Figure \ref{fig:MMVSim1TM10dB}. By comparing Figure \ref{fig:MMVSim1} and Figure \ref{fig:MMVSim1TM10dB}, we can observe obvious degradation in the LSM images, while there is no obvious degradation in the MMV images. Figure \ref{fig:MMVSim1_CV}(b) shows the residual curves, from which we can see the reconstruction residual of the optimal solution, $r_{\text{rec}}\approx0.105$, is higher than that of Figure \ref{fig:MMVSim1_CV}(a), $r_{\text{rec}}\approx0.025$.   

         Now let us process the TE-polarized data of different SNRs, 30 dB and 10 dB. Figure \ref{fig:MMVSim1TE} shows the scatterer shape reconstructed by MMV and LSM, respectively. The imaging results demonstrate again that, in the perspective of resolving ability, the proposed method outperforms the linear sampling method. In addition, the proposed method maintains good imaging performance for different SNRs. 

    \subsubsection{Simulation 2}

        \begin{figure}[!t]
            \centering
            \makebox[\columnwidth]{
            \subfloat[]{\includegraphics[height=0.375\linewidth] {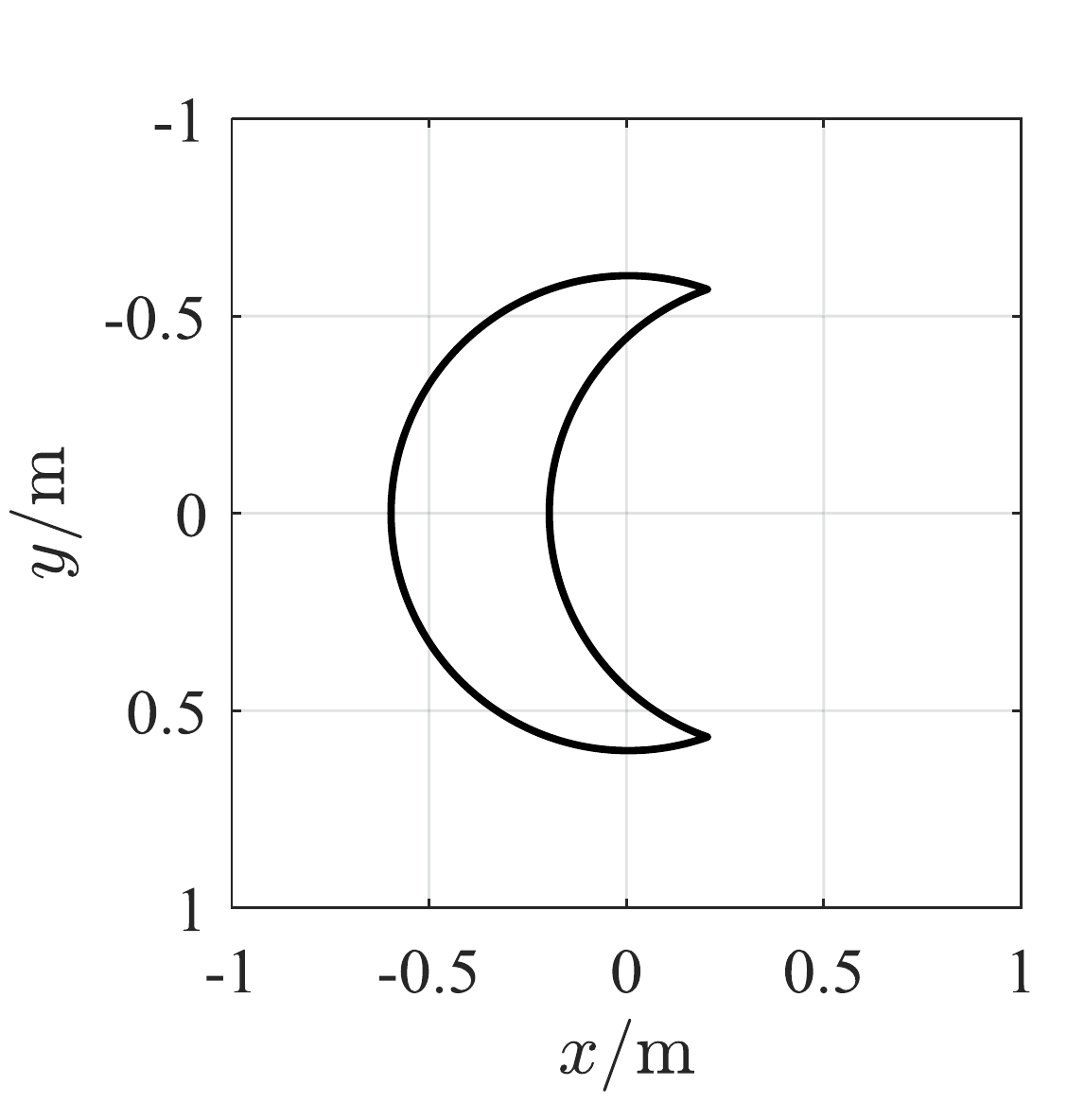}}\qquad
            \subfloat[]{\includegraphics[height=0.375\linewidth] {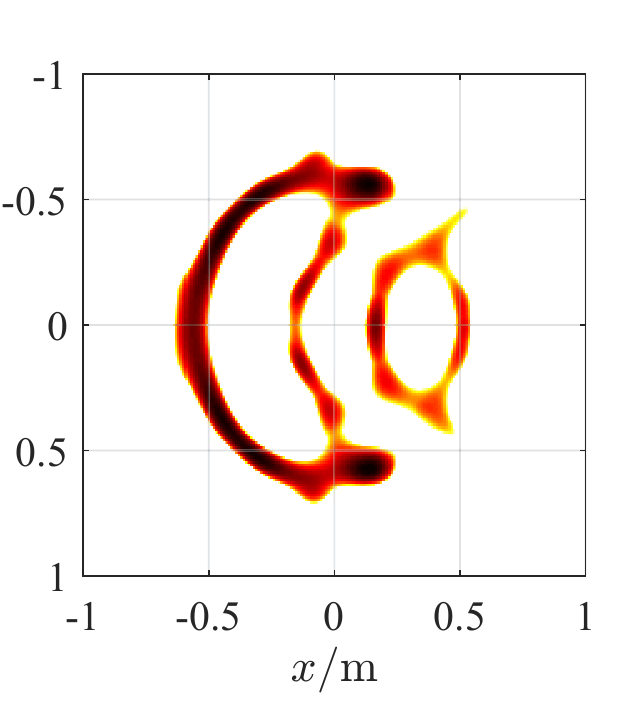}}} 
            \makebox[\columnwidth]{
            \subfloat[]{\includegraphics[height=0.375\linewidth] {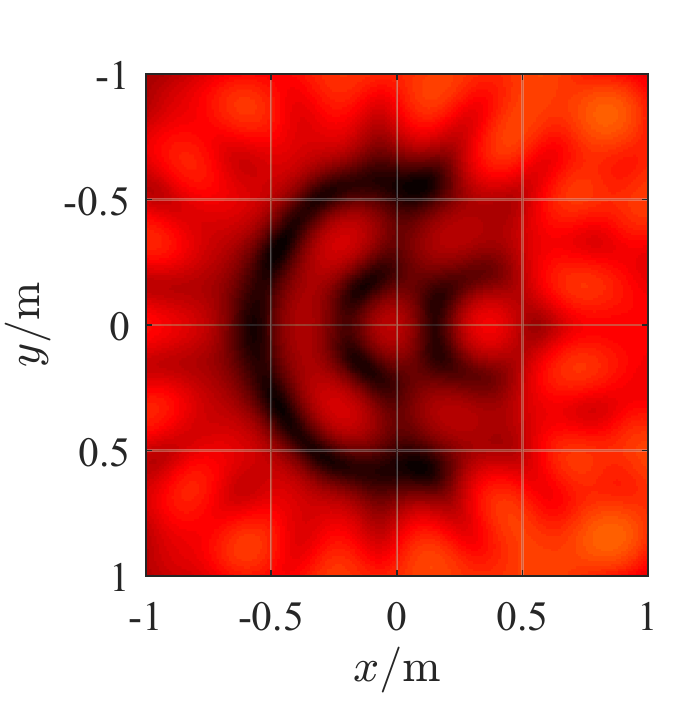}}\qquad
            \subfloat[]{\includegraphics[height=0.375\linewidth] {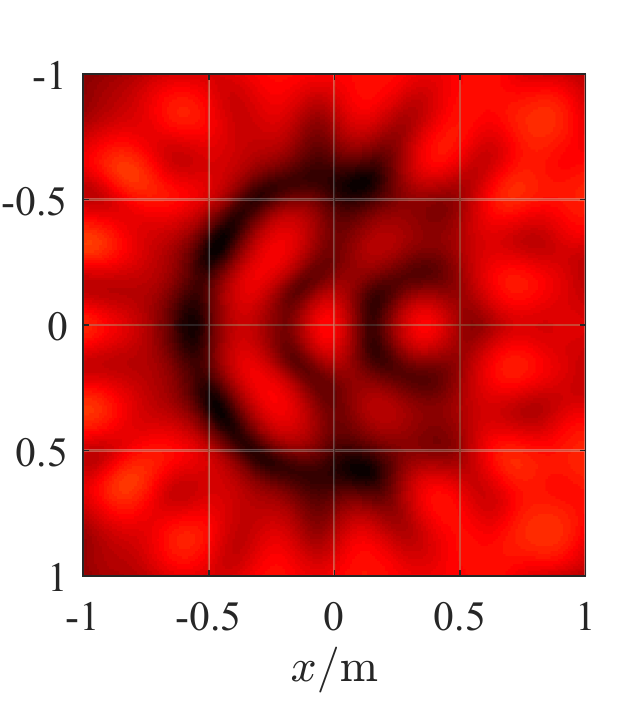}}}
            \raggedright
            \includegraphics[width=0.35\linewidth] {colorbar}
            \caption{Scatterer geometry and its reconstructed shapes in Simulation 2. 30 dB Gaussian noise is added to the measurement data. (a) Scatterer geometry; The scatterer shape (the value of the indicator function in dB) reconstructed by processing the TM-polarized data with MMV (b), LSM (c), and improved LSM with $I=6$ (d), respectively.}
            \label{fig:MMVSim2}
        \end{figure}

        \begin{figure}[!t]
            \centering
            \makebox[\columnwidth]{
            \subfloat[]{\includegraphics[height=0.375\linewidth] {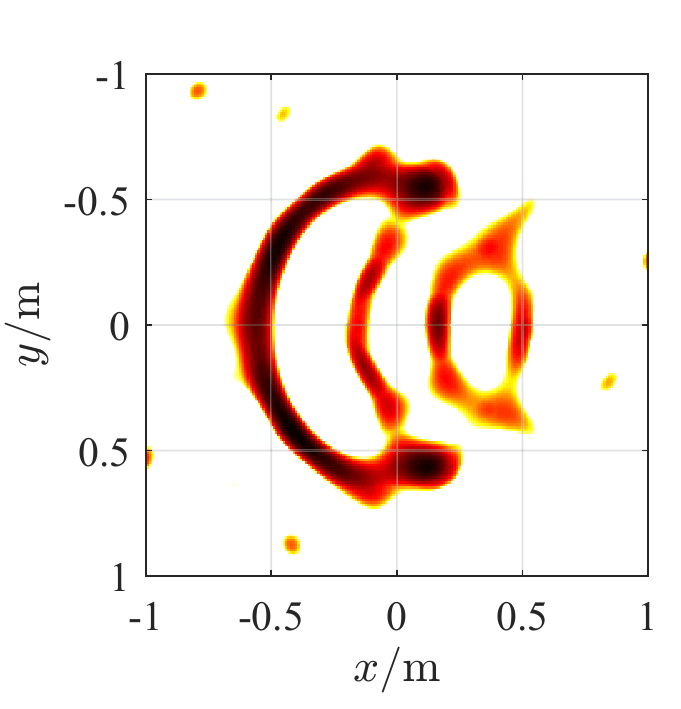}}} 
            \makebox[\columnwidth]{
            \subfloat[]{\includegraphics[height=0.375\linewidth] {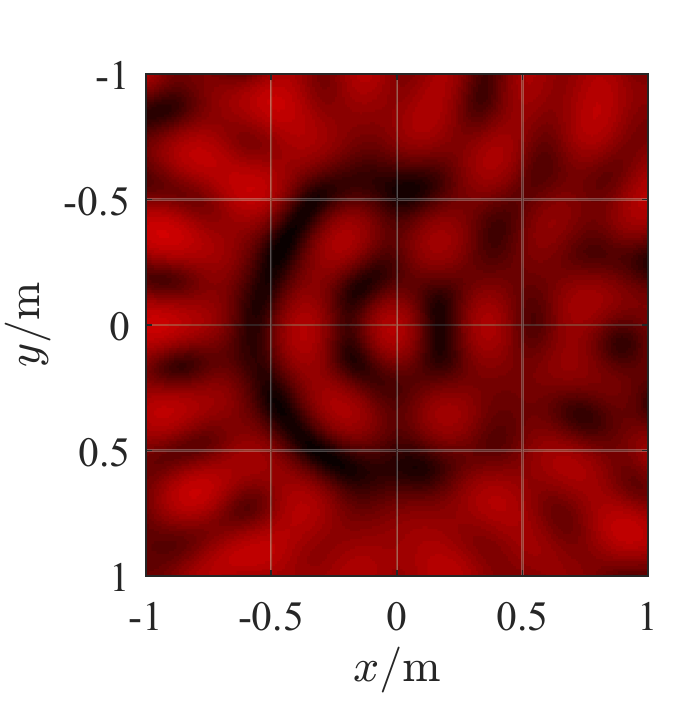}}\qquad
            \subfloat[]{\includegraphics[height=0.375\linewidth] {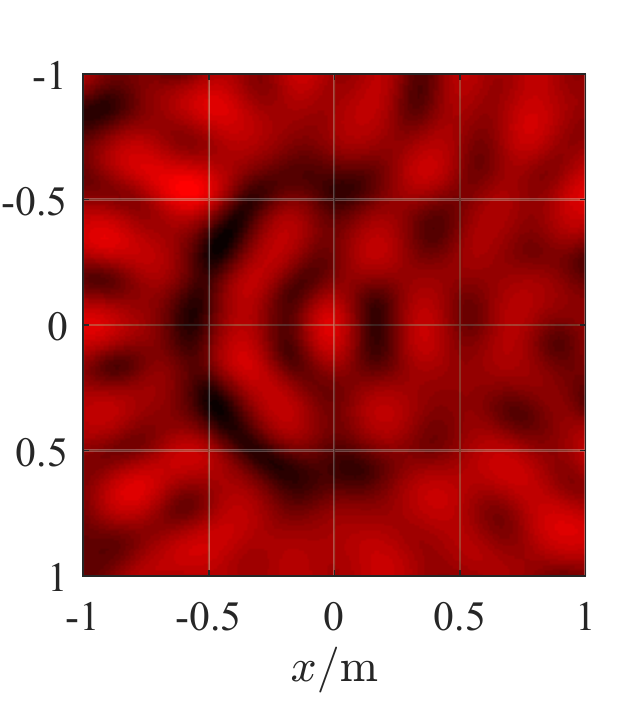}}}
            \raggedright
            \includegraphics[width=0.35\linewidth] {colorbar}
            \caption{Scatterer geometry and its reconstructed shapes in Simulation 2. 10 dB Gaussian noise is added to the measurement data. The scatterer shape (the value of the indicator function in dB) reconstructed by processing the TM-polarized data with MMV (a), LSM (b), and improved LSM with $I=7$ (c), respectively.}
            \label{fig:MMVSim2TM10dB}
        \end{figure}

        \begin{figure}[!t]
            \centering
            \makebox[\columnwidth]{
            \subfloat[]{\includegraphics[height=0.375\linewidth] {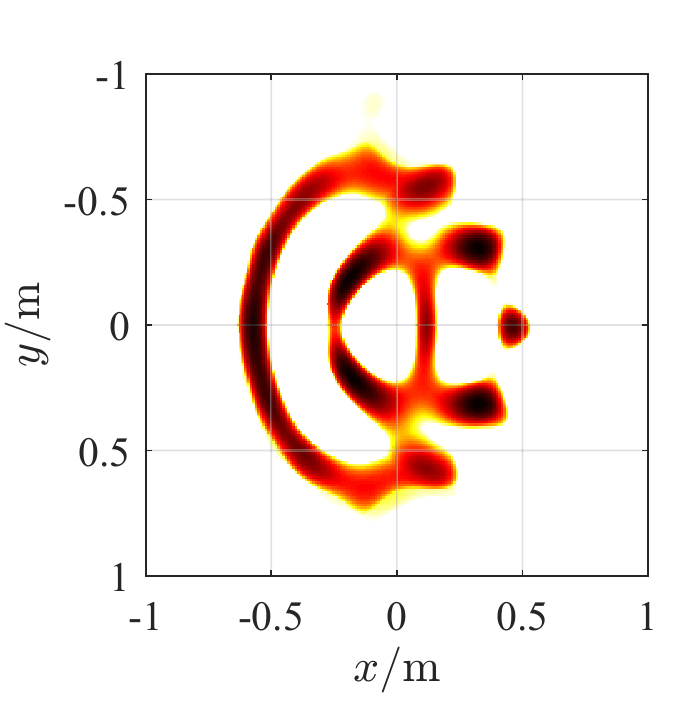}}\qquad
            \subfloat[]{\includegraphics[height=0.375\linewidth] {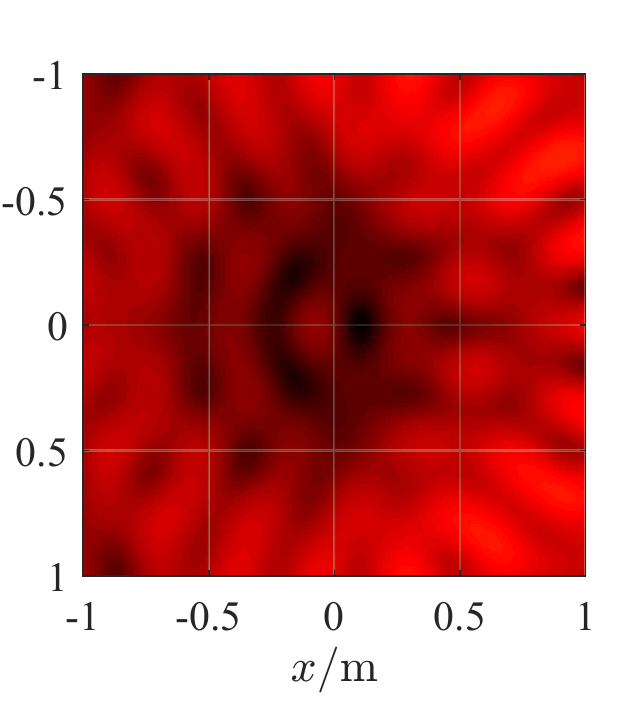}}}
            \makebox[\columnwidth]{
            \subfloat[]{\includegraphics[height=0.375\linewidth] {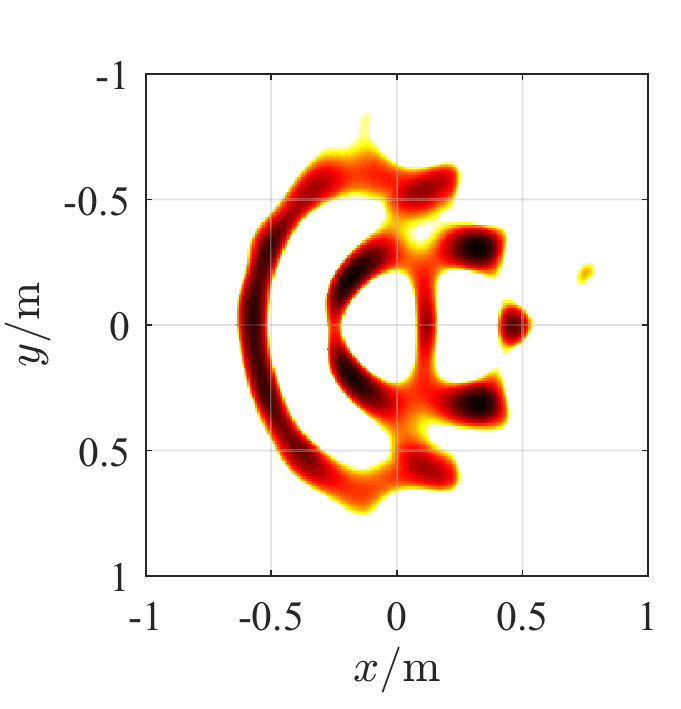}}\qquad
            \subfloat[]{\includegraphics[height=0.375\linewidth] {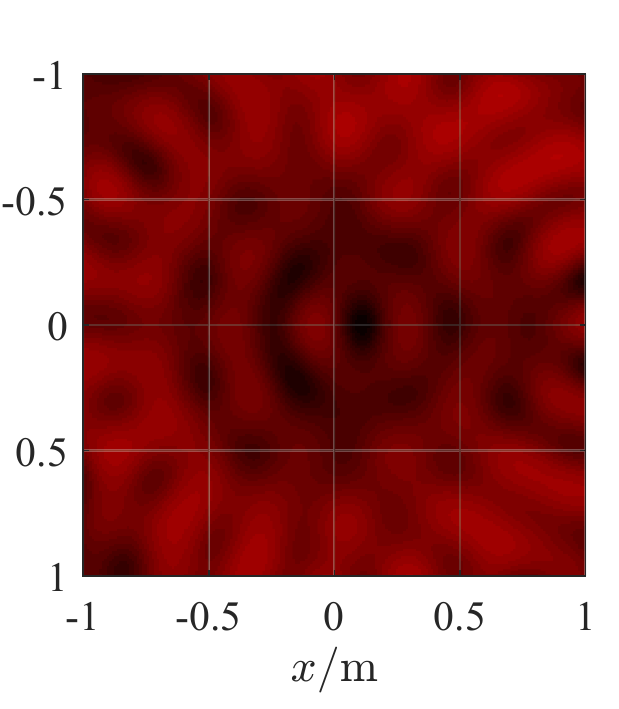}}}
            \raggedright
            \includegraphics[width=0.35\linewidth] {colorbar}
            \caption{Scatterer shape (the value of the indicator function in dB) reconstructed by processing the TE-polarized data of Simulation 2. (a), (c): MMV images; (b), (d): LSM images; (a), (b): SNR$=30$ dB; (c), (d): SNR$=10$ dB.}
            \label{fig:MMVSim2TE}
        \end{figure}

        In the second simulation, we restrict the inversion domain to $[-1.0,1.0] \times [-1.0,1.0]$ m$^2$, in which the target is fully covered. The inversion domain is discretized with equal grid size of $0.01$ m ($=\lambda/60$). First, we process the TM-polarized data of 30 SNR by MMV and LSM, respectively. Figure \ref{fig:MMVSim2}(b), (c), and (d) show the reconstructed shape by MMV, LSM, and improved LSM, respectively. We can see from the results that the boundary at the left side is well reconstructed by the three methods, while the arch at the right side shows more artifacts in both Figure \ref{fig:MMVSim2}(b) and (c), because the arch at the right side is concave and multi-path scattering is more severe than the left side which is convex. Comparison of Figure \ref{fig:MMVSim2}(c) and (d) shows minor suppression of artifacts in the interior of the cylinder by improved LSM. The imaging results of 10 dB SNR data are shown in Figure \ref{fig:MMVSim2TM10dB}. Apart from some minor artifacts, no obvious degradation occurs in MMV image, while we can observe severe degradation of image resolution in LSM images.  

        The MMV image and LSM image obtained by processing the TE-polarized data of 30 dB SNR and 10 dB SNR are shown in Figure \ref{fig:MMVSim2TE}. From the results, we can observe that the proposed method is able to reconstruction the scatterer's shape with some artifacts occur at the concave side, while LSM fail to give the basic profile of the target. Considering the length of this paper, the residual curves in this simulation are not given.

    \subsection{Analysis of computational complexity}

        \begin{table}[!t]
            \renewcommand{\arraystretch}{1.0}
            \caption{Running times of the two numerical examples}
            \label{tab.MMV_Num_Time}
            \centering \makebox[\columnwidth]{
            \begin{tabular}{|c|>{\centering\arraybackslash}m{1.0 cm}|>{\centering\arraybackslash}m{1.4 cm}|>{\centering\arraybackslash}m{1.4 cm}|>{\centering\arraybackslash}m{1.3 cm}|}
            \hline
            Example    &   Pol.    &   MMV [s]    &   LSM [s] &  Improved LSM [s]\\ 
            \hline
            \multirow{2}{*}{1}      &  TM  &   14.7     &   0.0286    &   0.2711    \\ \cline{2-5}
                                    &  TE  &   45.3     &   0.0599    &   \slash    \\ \hline
            \multirow{2}{*}{2}      &  TM  &   11.2     &   0.0290    &   0.2429    \\ \cline{2-5}
                                    &  TE  &   38.8     &   0.0587    &   \slash    \\ \hline
            \end{tabular}
            }
        \end{table}

        The sensing matrix $\bm\Phi$ can be computed (or analytically given for the experiments in homogeneous background) and stored beforehand. It is easy to see from Algorithm \ref{alg.SPGTM} and Algorithm \ref{alg.NewtonRootTM} that, the computational complexity of the GMMV-based linear method primarily depends on the number of matrix-vector multiplications, $\bm\Phi\bm{J}$ and $\bm{\Phi}^H\bm{R}$. Empirically, each iteration involves maximally 2 times of $\bm\Phi\bm{J}$ and 1 time of $\bm{\Phi}^H\bm{R}$. In order to study the computational complexity of the proposed algorithms, we use one complex data multiplication as the measurement unit. The computational complexity for computing $\bm\Phi\bm{J}$ and $\bm{\Phi}^H\bm{R}$ is $QN$ for the TM case and $4QN$ for the TE case. Let us use $L$ to denote the iteration number, then the computational complexity of the proposed method in the TM and TE case is
        \begin{subequations}
          \begin{align}
            C_{\text{TM}}&=3LQN,\\
            C_{\text{TE}}&=12LQN,
          \end{align}
        \end{subequations}
        respectively. When the mesh gets finer or the inversion domain gets larger, the iteration number, $L$, almost keeps unchanged, and the running time therefore increases linearly with the grid number.

        In our experiments, the imaging algorithms are implemented with MATLAB language. We ran the program on a desktop with one Intel(R) Core(TM) i5-3470 CPU @ 3.20 GHz, and we did not use parallel computing. Table~\ref{tab.MMV_Num_Time} lists the running times of the proposed method, LSM, and improved LSM in the two simulations. As one can see that the computation time of the proposed method is hundreds of times longer than that of LSM and tens of times longer than improved LSM. The most time-consuming part of the proposed method is the matrix-vector multiplication in each iteration, while LSM only calls singular value decomposition to the measurement data matrix for once. However, the running times of the proposed method are still acceptable in view of the gain of resolving ability.

\section{Experimental Data Imaging}\label{sec.Exp}

    \begin{figure}[!t]
        \centering
        \includegraphics[width=0.45\linewidth]{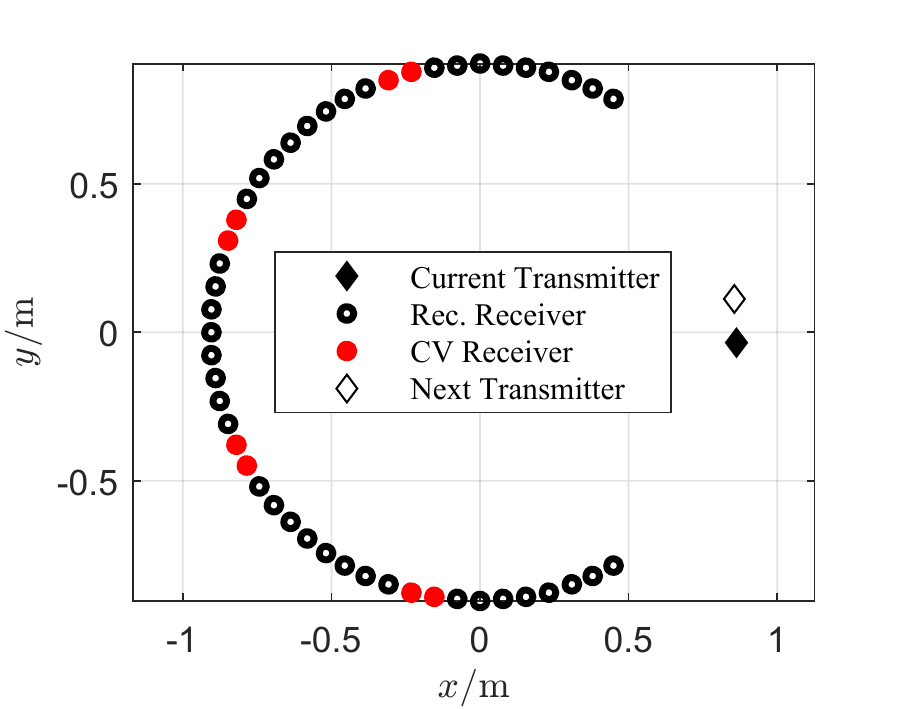}
        \caption{Measurement configuration of the Fresnel data-sets: \textit{rectTM\_cent}, \textit{uTM\_shaped}, and \textit{rectTE\_8f}.}
        \label{fig:MMV_FresnelI_Conf}
    \end{figure}

    \begin{figure}[!t]
        \centering
        \subfloat[]{\includegraphics[width=0.35\linewidth] {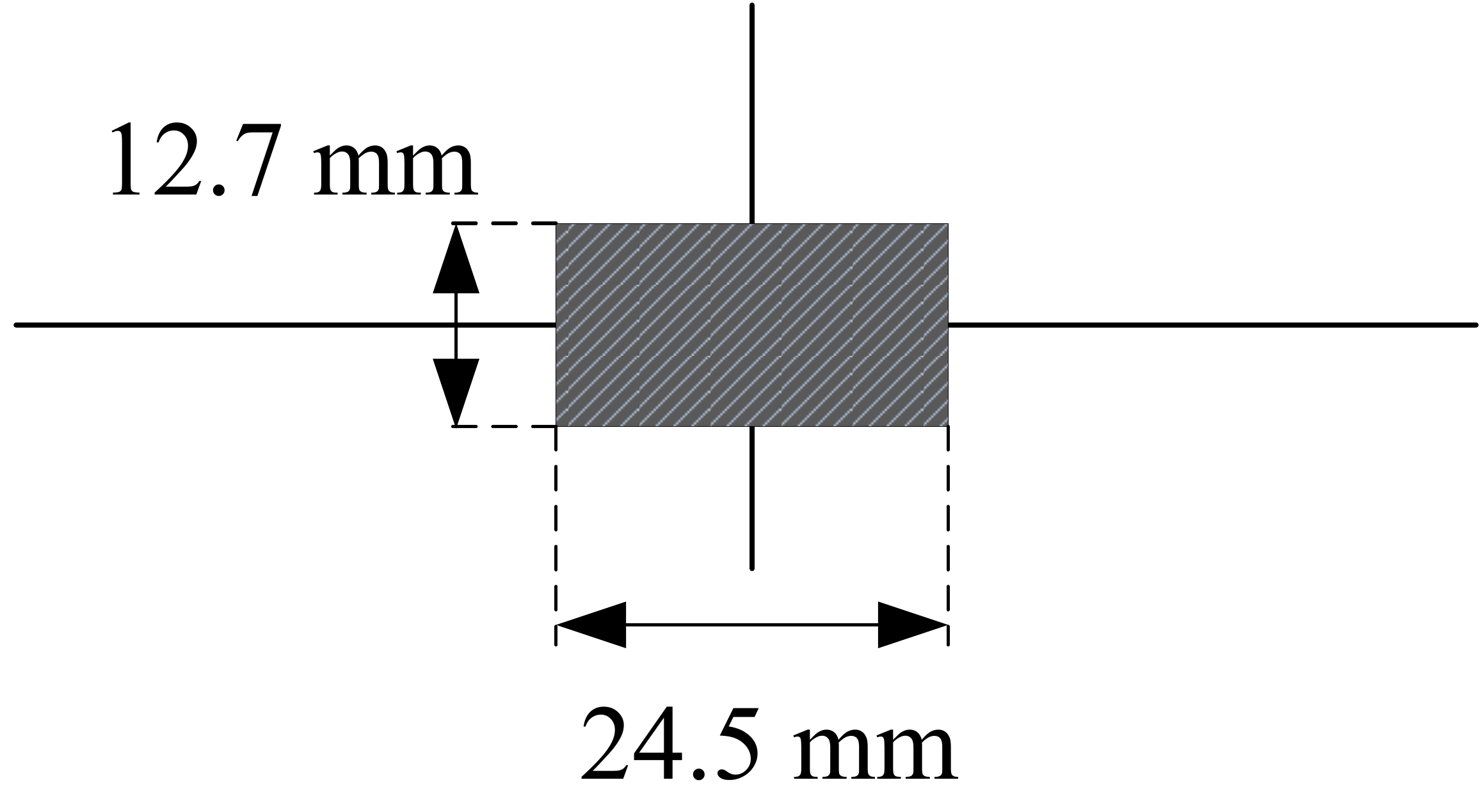}}\qquad
        \subfloat[]{\includegraphics[width=0.35\linewidth] {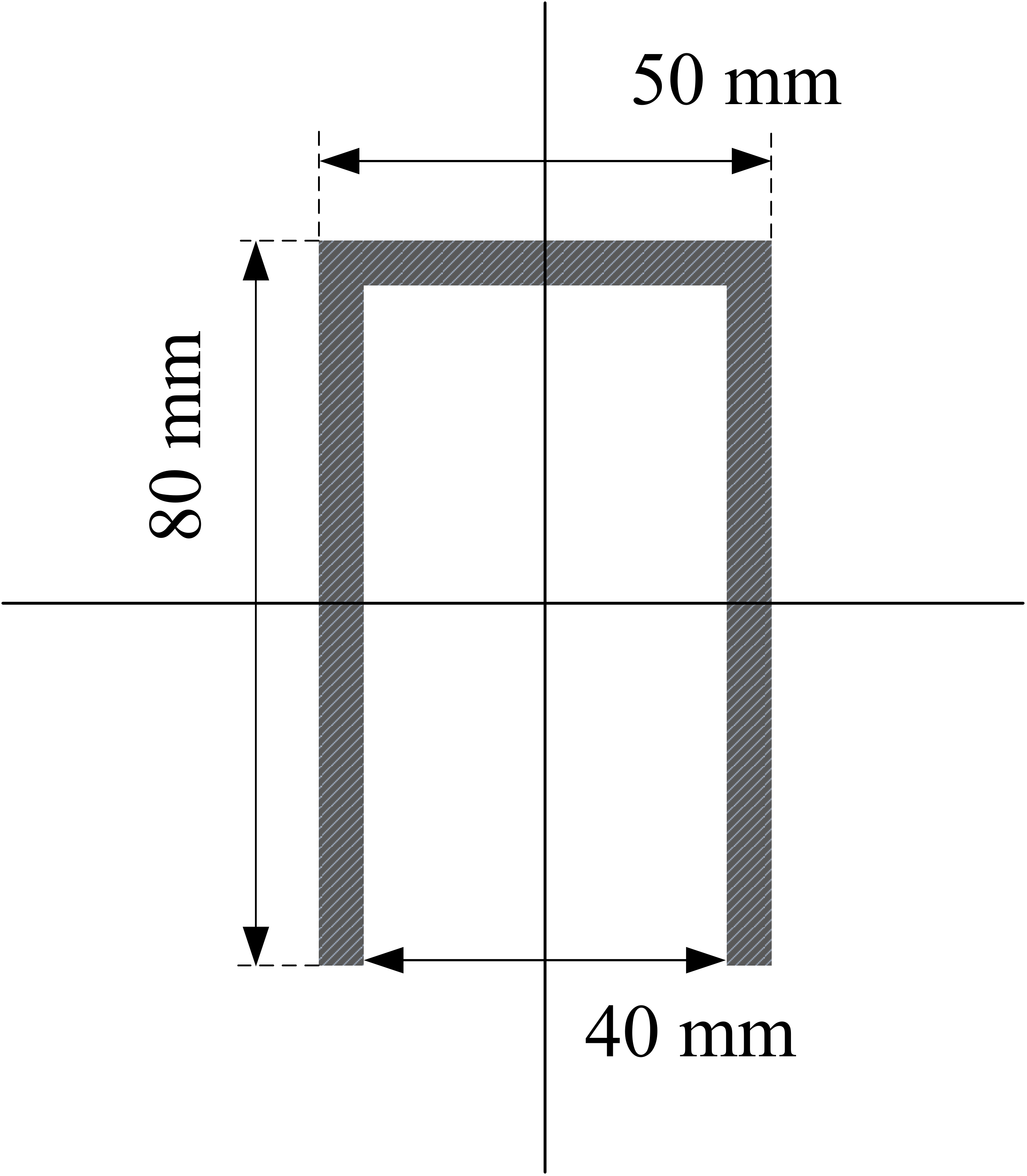}}\\
        \caption{Geometry of the scatterers: (a) The rectangular metallic cylinder; (b) The ``U-shaped'' metallic cylinder.}
        \label{fig:FreIobj}
    \end{figure}

    \begin{figure}[!t]
        \centering
        \subfloat[]{\includegraphics[height=0.375\linewidth] {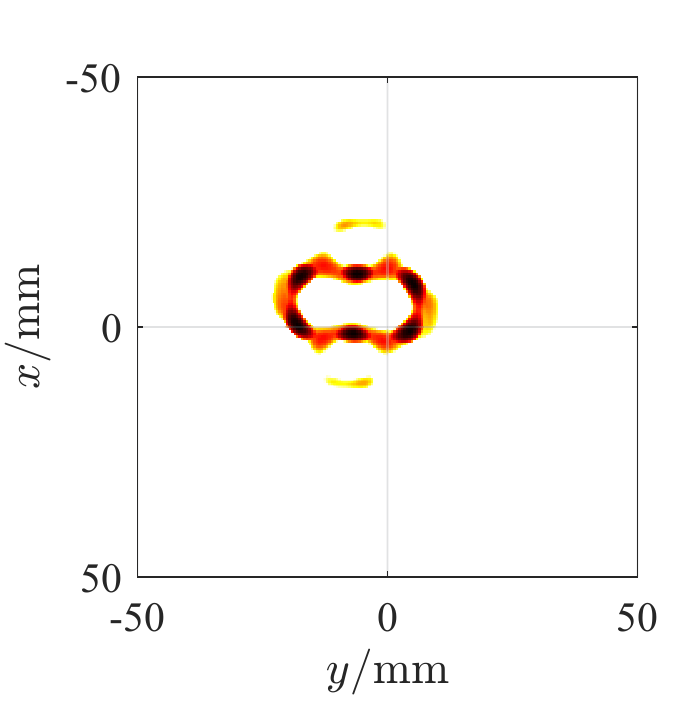}}\\ 
        \makebox[\columnwidth]{
        \subfloat[]{\includegraphics[height=0.375\linewidth] {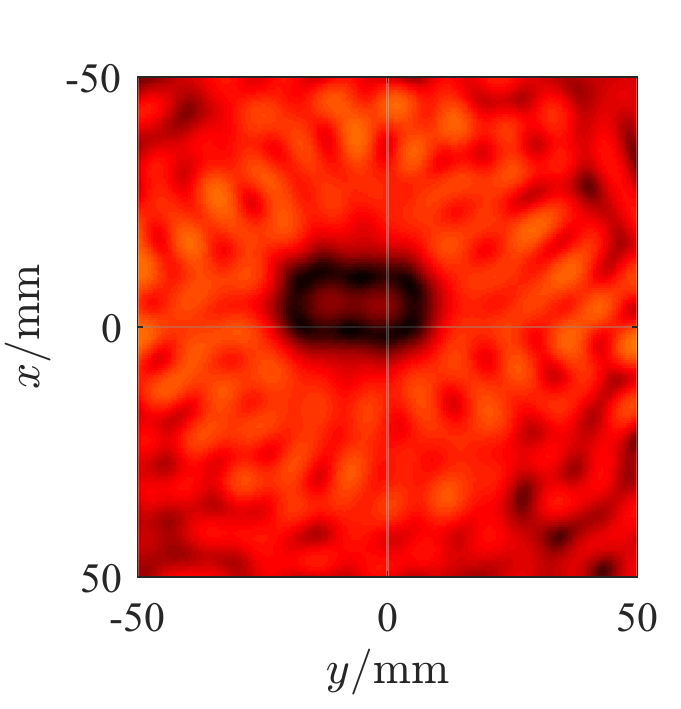}}\qquad
        \subfloat[]{\includegraphics[height=0.375\linewidth] {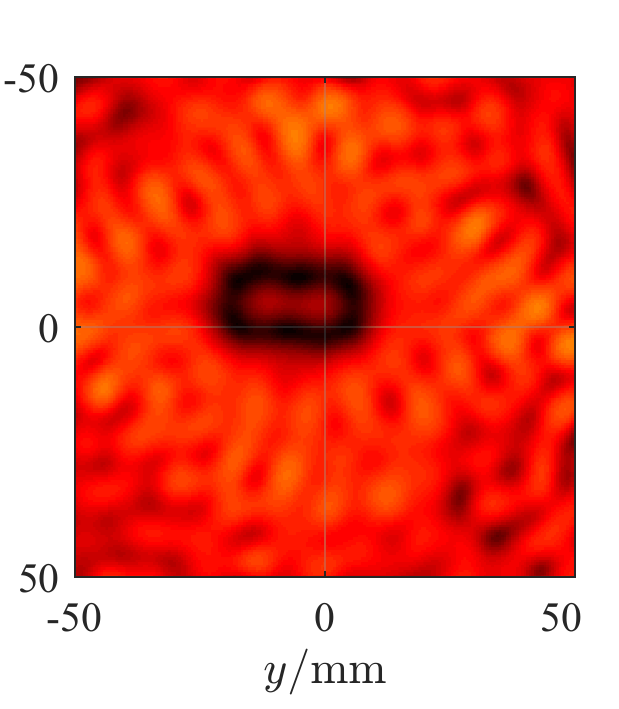}}}
        \raggedright
        \includegraphics[width=0.35\linewidth] {colorbar}
        \caption{Scatterer shape (the value of the indicator function in dB) reconstructed by processing the TM-polarized data-set: \textit{rectTM\_cent} at 16 GHz with MMV (a), LSM (b), and improved LSM with $I=9$ (c), respectively. 18 transmitter positions and 49 receiver positions for each transmitter are selected for imaging.}
        \label{fig:MMVrectTM_cent16GHz}
    \end{figure}

    \begin{figure}[!t]
        \centering
        \subfloat[]{\includegraphics[height=0.375\linewidth] {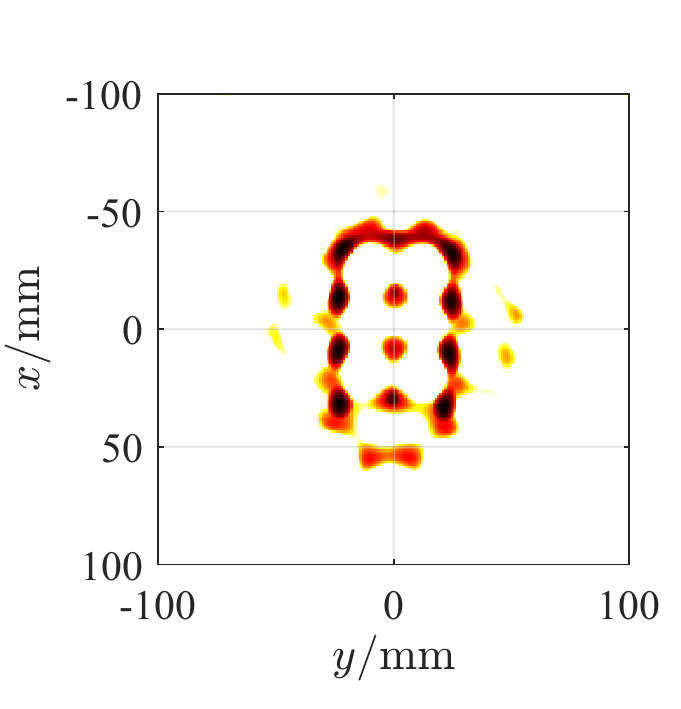}}\\ 
        \makebox[\columnwidth]{
        \subfloat[]{\includegraphics[height=0.375\linewidth] {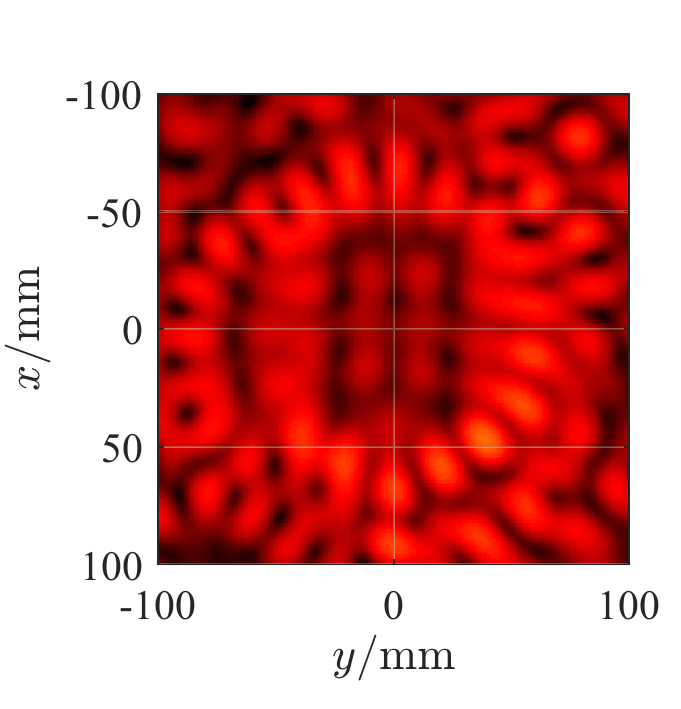}}\qquad
        \subfloat[]{\includegraphics[height=0.375\linewidth] {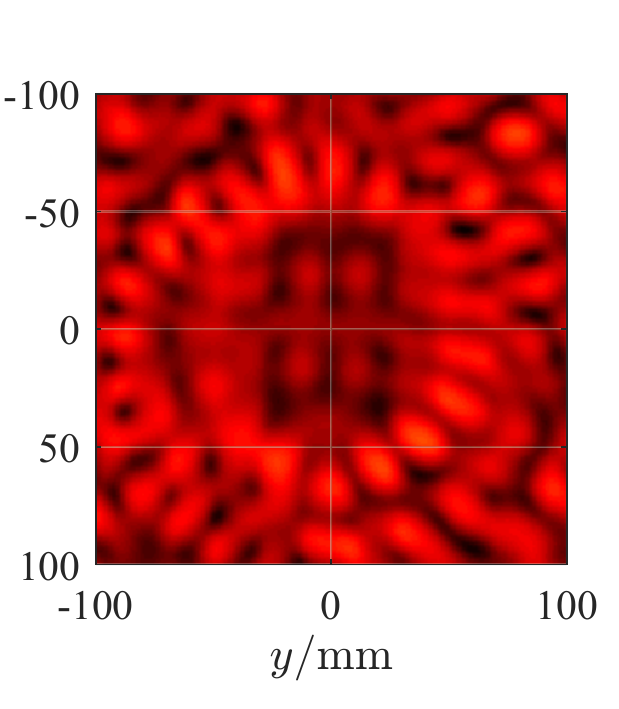}}}
        \raggedright
        \includegraphics[width=0.35\linewidth] {colorbar}
        \caption{Scatterer shape (the value of the indicator function in dB) reconstructed by processing the TM-polarized data-set: \textit{uTM\_shaped} at 8 GHz with MMV (a), LSM (b), and improved LSM with $I=8$ (c), respectively. 18 transmitter positions and 49 receiver positions for each transmitter are selected for imaging.}
        \label{fig:MMVuTM_shaped8GHz}
    \end{figure}
 
    \begin{figure}[!t]
        \centering
        \makebox[\columnwidth]{
        \subfloat[]{\includegraphics[height=0.375\linewidth] {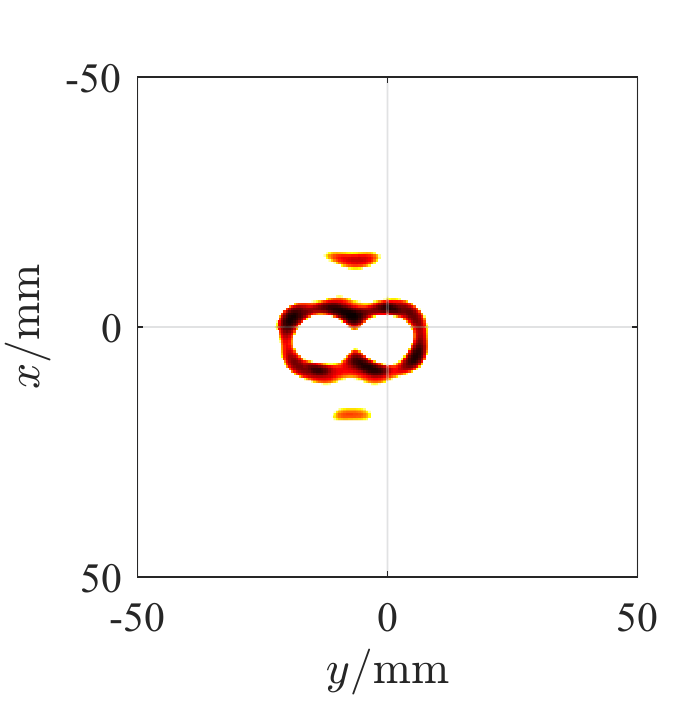}}\qquad
        \subfloat[]{\includegraphics[height=0.375\linewidth] {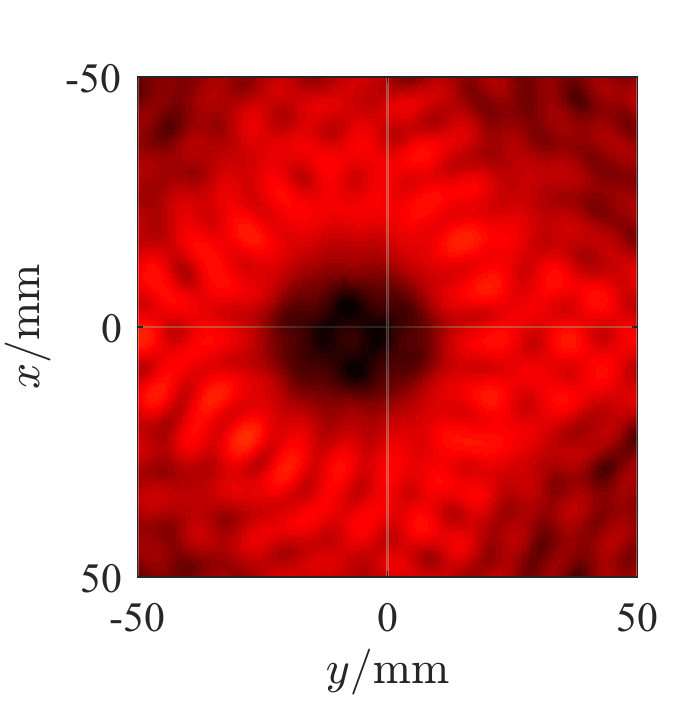}}}
        \raggedright
        \includegraphics[width=0.35\linewidth] {colorbar}
        \caption{Scatterer shape (the value of the indicator function in dB) reconstructed by processing the TE-polarized data-set: \textit{rectTE\_8f} at 16 GHz with MMV (a) and LSM (b), respectively. 18 transmitter positions and 49 receiver positions for each transmitter are selected for imaging.}
        \label{fig:rectTE_8f16GHz}
    \end{figure}

    \begin{figure}[!t]
        \centering
        \subfloat[]{\includegraphics[width=0.475\linewidth] {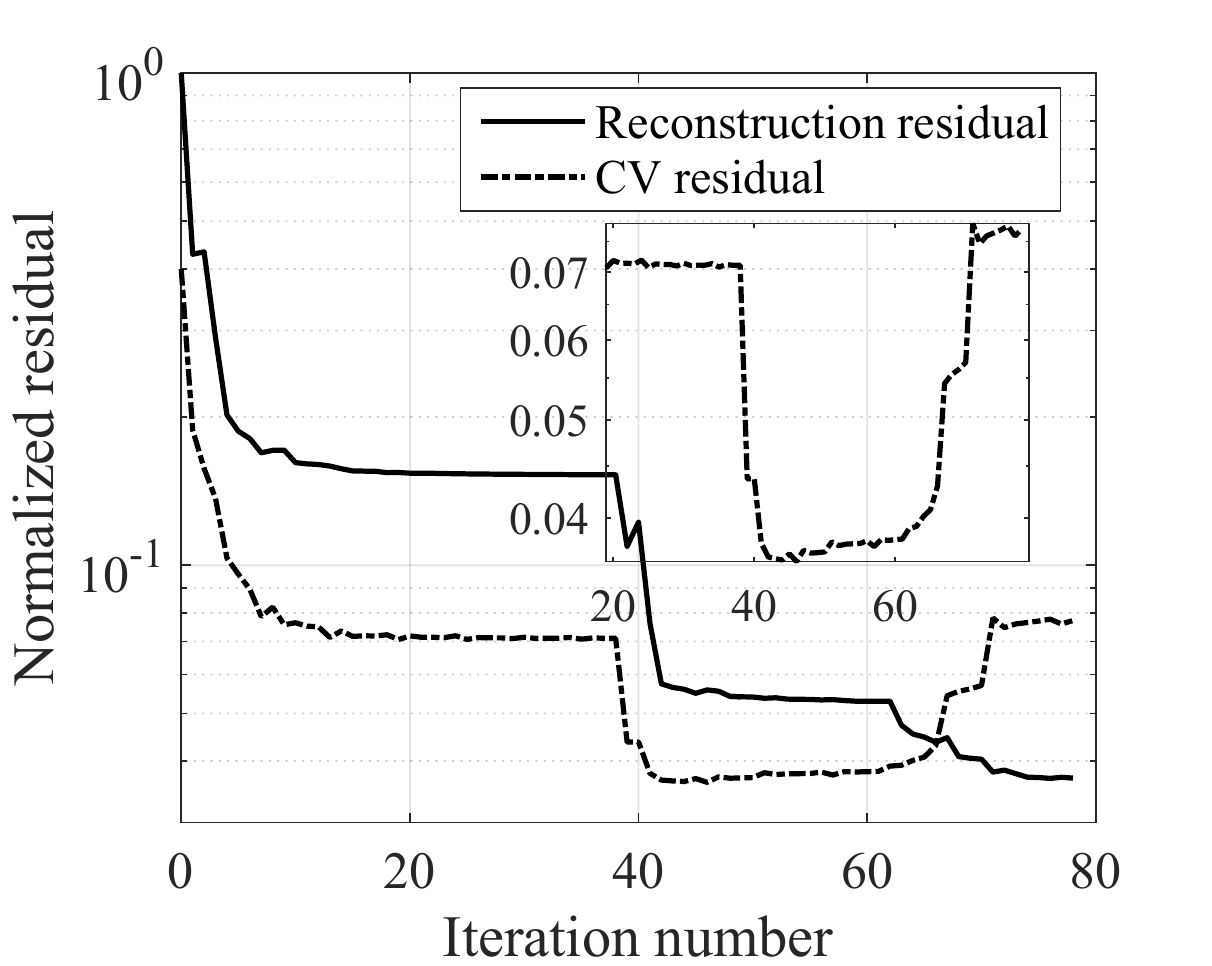}}\\
        \subfloat[]{\includegraphics[width=0.475\linewidth] {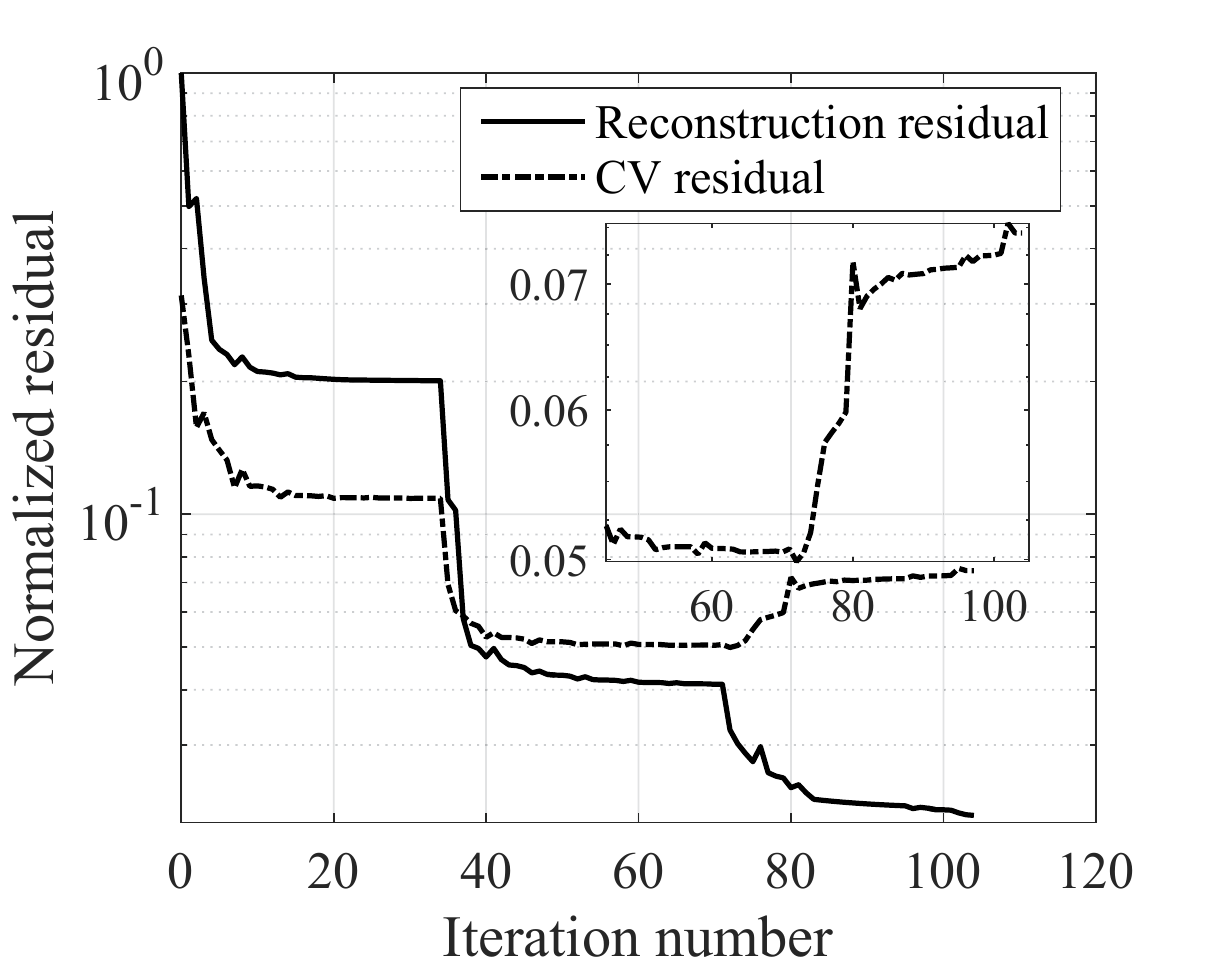}}
        \subfloat[]{\includegraphics[width=0.475\linewidth] {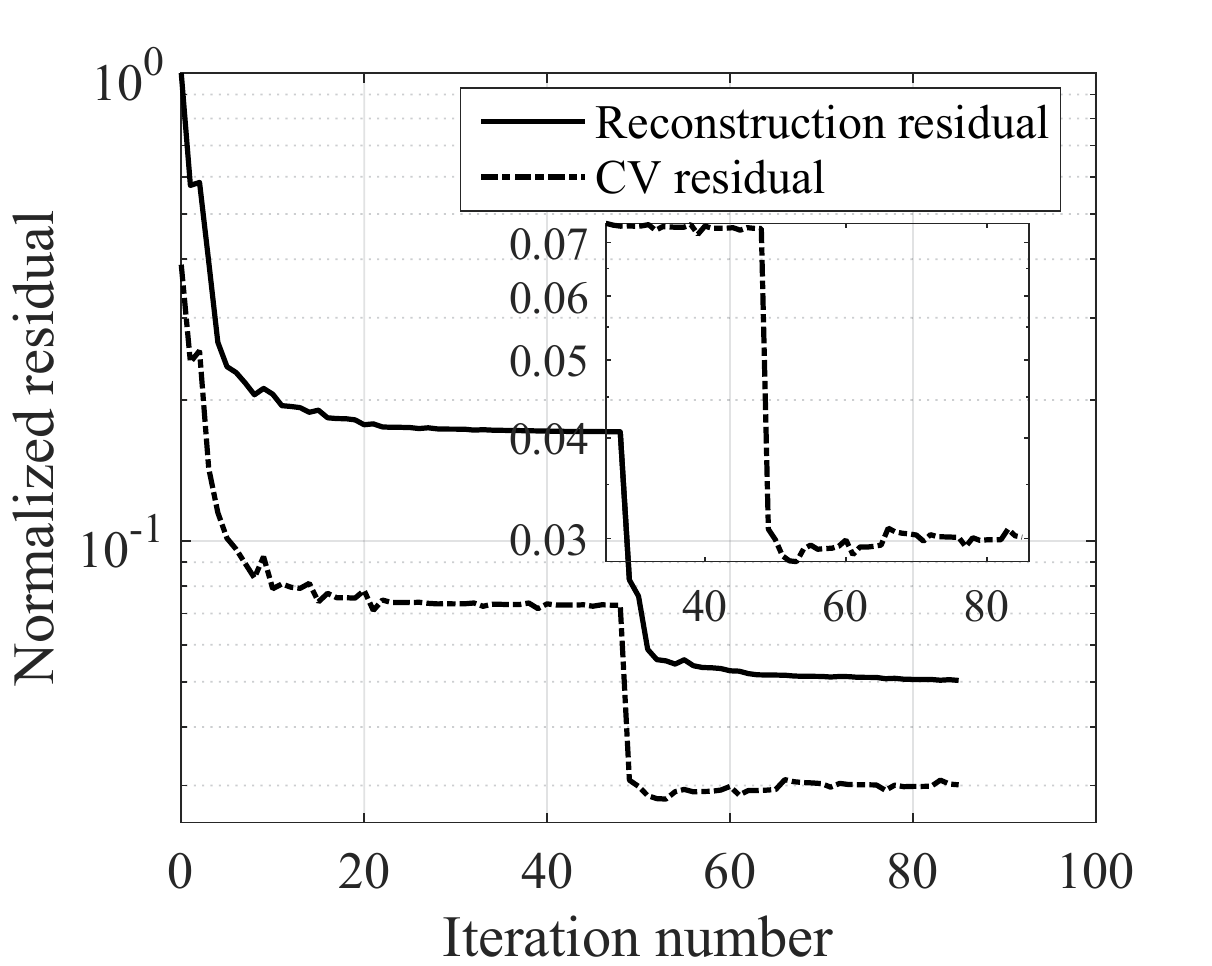}} 
        \caption{Reconstruction residual curve and CV residual curve of the Fresnel data-sets: \textit{rectTM\_cent} (a), \textit{uTM\_shaped} (b), and \textit{rectTE\_8f} (c), at 16 GHz, 8 GHz, and 16 GHz, respectively.}
        \label{fig:MMV_FresnelI_CV}
    \end{figure}

    \begin{table}[!t]
        \renewcommand{\arraystretch}{1.0}
        \caption{Running times of the experimental examples}
        \label{tab.MMV_Exp_Time}
        \centering \makebox[\columnwidth]{
        \begin{tabular}{|c|>{\centering\arraybackslash}m{1.6 cm}|>{\centering\arraybackslash}m{1.4 cm}|>{\centering\arraybackslash}m{1.4 cm}|>{\centering\arraybackslash}m{1.3 cm}|}
        \hline
        Data-set                  &   MMV [s]     &   LSM [s]     &  Improved LSM [s] \\ \hline
        \textit{rectTM\_cent}     &   9.2         &   0.0312      &   0.3486          \\ \hline
        \textit{uTM\_shaped}      &   12.4        &   0.0277      &   0.3081          \\ \hline
        \textit{rectTE\_8f}       &   32.9        &   0.0676      &   \slash          \\ \hline
        \end{tabular}
        }
    \end{table}

    In this subsection, we applied our method to the experimental data provided by the Remote Sensing and Microwave Experiments Team at the Institut Fresnel, France, using an HP8530 network analyzer \cite{0266-5611-17-6-301}. The experimental setup consists of a large anechoic chamber, $14.50$ m long, $6.50$ m wide and $6.50$ m high, with a set of three positioners to adjust antennas or target positions. A 2-D bistatic measurement system is considered, with an emitter placed at a fixed position on the circular rail, while a receiver is rotating with the arm around a vertical cylindrical target. The targets rotated from 0$^{\circ}$ to 350$^{\circ}$ in steps of 10$^{\circ}$ with a radius of $720\pm3$ mm, and the receiver rotated from 60$^{\circ}$ to 300$^{\circ}$ in steps of 5$^{\circ}$ with a radius of $760\pm3$ mm. In TE polarization, only the component orthogonal to both the invariance axis of the cylinder and the direction of illumination is measured. Figure \ref{fig:MMV_FresnelI_Conf} gives the measurement configuration, in which the selection of CV measurements and reconstruction measurements is illustrated. The time dependence in this experiment is $\exp(\text{i}\omega t)$. Therefore, after subtracting the incident field from the total field, the measurement data can be directly used for inversion. The targets we consider here are a rectangular metallic cylinder and a ``U-shaped'' metallic cylinder, which have been shown by Figure \ref{fig:FreIobj}(a) and (b). Three data-sets are processed: \textit{rectTM\_cent} at 16 GHz, \textit{uTM\_shaped} at 8 GHz, and \textit{rectTE\_8f} at 16 GHz. The selected frequencies correspond to wavelengths which are comparable to the dimension of the targets, i.e., the resonance frequency range.

    As we have discussed in Subsection \ref{subsec.det.mea.conf}, 18 transmitters of $20^{\circ}$ arc interval are enough for the proposed method to give good resolution. Therefore, in the following experiments, only 18 transmitter positions (half of the measurement data) are used for imaging. First, let us process the TM-polarized data-set: \textit{rectTM\_cent} at 16 GHz. The inversion domain for imaging the rectangular metallic cylinder is restricted to $[-50,50]\times[-50,50]$ mm$^2$, and the inversion domain is discretized with equal grid size of $0.5$ mm ($=\lambda/37.5$). Figure~\ref{fig:MMVrectTM_cent16GHz}(a), (b), and (c) show the scatterer shape reconstructed by MMV, LSM, and improved LSM, respectively. The residual curves are shown in Figure \ref{fig:MMV_FresnelI_CV}(a). From the imaging results we observe that Figure~\ref{fig:MMVrectTM_cent16GHz}(a) shows higher resolution and less artifacts than Figure~\ref{fig:MMVrectTM_cent16GHz}(b) and (c). And we also observe that there is no big difference between Figure~\ref{fig:MMVrectTM_cent16GHz}(b) and (c).

    Now let us consider the ``U-shaped'' metallic cylinder. The inversion domain is restricted to $[-100,100]\times[-100,100]$ mm$^2$, and the inversion domain is discretized with equal grid size of $1$ mm ($=\lambda/37.5$). Figure \ref{fig:MMVuTM_shaped8GHz}(a), (b), and (c) give the scatterer shape reconstructed by MMV, LSM, and improved LSM, respectively. The residual curves are shown in Figure \ref{fig:MMV_FresnelI_CV}(b). Severe artifacts can be observed in LSM image and improved LSM image. Furthermore, the suppression to the artifacts is not obvious in the improved LSM image. In the contrary, the ``U-shaped'' cylinder is reconstructed by the proposed method with the boundary well distinguished. Some artifacts can be observed vertically aligned in the interior and below the opening which are caused by the complicated scattering in the opening area. 

    Finally, let us process the TE-polarized data-set: \textit{rectTE\_8f} at 16 GHz. The scatterer shape reconstructed by MMV and LSM is shown in Figure \ref{fig:rectTE_8f16GHz}(a) and (b), respectively. The residual curves are shown in Figure \ref{fig:MMV_FresnelI_CV}(c). It can be observed that the boundary of the rectangular metallic cylinder is not distinguishable in Figure \ref{fig:rectTE_8f16GHz}(b), while in Figure \ref{fig:rectTE_8f16GHz}(a) the rectangular boundary can be clearly distinguished. The data processing is performed on the same computing platform, and the running times of all the methods are listed in Table~\ref{tab.MMV_Exp_Time}. In the end, we summarize this section as follows: 1) the proposed method is able to obtain higher resolution than LSM and improved LSM; 2) the proposed method is more computationally expensive than LSM and improved LSM; 3) the suppression to the artifacts by improved LSM is not obvious when less transmitters are used.

\section{Conclusion}\label{sec.Con}

    In this paper, we addressed the nonlinear inverse scattering problem of highly conductive objects with a linear model. Multiple measurement vectors model is exploited and sum-of-norm constraint is used as a regularization approach. A CV-based modified SPGL1 method is proposed to invert the data without estimating the noise level. Numerical results and experimental results of both TM polarization and TE polarization demonstrate that the proposed method shows higher resolving ability than both LSM and improved LSM. In some cases where the latter two methods fail, the proposed method can still successfully recover the profile of the targets. Numerical experiments also demonstrate that the MMV method maintains good imaging performance in the disturbance of 10 dB SNR Gaussian random noise. The running time of the proposed method is hundreds of times longer than LSM and tens of times longer than improved LSM. However, it is still promising in view of the gain of resolving performance and the linear relationship between the computational complexity and the size of the inversion domain. In addition, it might fail in the presence of not conductive scatterers, which is an obvious limitation of the proposed method.

\bibliographystyle{IEEEtran}
\bibliography{mybib}

\end{document}